\documentclass[sigconf]{acmart}
\AtBeginDocument{%
  }
\usepackage{tikz}

\newcommand{\bc}[1]{%
  \tikz[baseline=(char.base)]{
    \node[shape=circle, fill=black, text=white, inner sep=1pt] (char) {#1};
  }%
}
\usepackage{amsmath,amsfonts}
\usepackage{lipsum}
\usepackage{soul}
\usepackage{adjustbox}
\usepackage{algorithmic}
\usepackage{booktabs}
\usepackage{graphicx}
\usepackage{subcaption}
\usepackage{textcomp}
\usepackage{todonotes}
\usepackage{cleveref}
\usepackage[most]{tcolorbox}
\usepackage[inline]{enumitem}
\usepackage{multirow}
\usepackage{makecell}
\usepackage{algorithmic}
\usepackage{stfloats}
\usepackage{mwe}
\usepackage[font=normalsize]{caption}
\usepackage[font=bf]{caption}
\AtBeginDocument{%
  }
\usepackage{listings}
\usepackage{url}
\usepackage[normalem]{ulem}
\useunder{\uline}{\ul}{}
\usepackage{xcolor}
\usepackage{hyperref}
\definecolor{darkgreen}{RGB}{0,128,0} 
\definecolor{deepblue}{RGB}{0,19,222}
\hypersetup{
    colorlinks = true,
    urlcolor  = deepblue,
    citecolor = deepblue,
    linkcolor = deepblue
}
\lstdefinestyle{mystyle}{
    backgroundcolor=\color{backcolour},   
    commentstyle=\color{codegreen},
    keywordstyle=\color{magenta},
    numberstyle=\tiny\color{codegray},
    stringstyle=\color{codepurple},
    basicstyle=\ttfamily\footnotesize,
    breakatwhitespace=true,         
    breaklines=false,                 
    captionpos=b,                    
    keepspaces=true,                 
    numbers=left,                    
    numbersep=5pt,                  
    showspaces=false,                
    showstringspaces=false,
    showtabs=false,                  
    tabsize=2
}
\definecolor{codegreen}{rgb}{0,0.6,0}
\definecolor{codegray}{rgb}{0.5,0.5,0.5}
\definecolor{codepurple}{rgb}{0.58,0,0.82}
\definecolor{backcolour}{rgb}{0.95,0.95,0.92}
\copyrightyear{2025}
\acmYear{2025}
\setcopyright{cc}
\setcctype{by-nc}
\acmConference[ICPE Companion '25]{Companion of the 16th ACM/SPEC International Conference on Performance Engineering}{May 5--9, 2025}{Toronto, ON, Canada}
\acmBooktitle{Companion of the 16th ACM/SPEC International Conference on Performance Engineering (ICPE Companion '25), May 5--9, 2025, Toronto, ON, Canada}
\acmDOI{10.1145/3680256.3721320}
\acmISBN{979-8-4007-1130-5/2025/05}

\begin{document}

\title{FAILS: A Framework for Automated Collection and Analysis \\ of LLM Service Incidents}

% \author{S\'andor Battaglini-Fischer$^{*}$, Nishanthi Srinivasan$^{*}$, B\'alint L\'aszl\'o Szarvas$^{*}$, \\ Xiaoyu Chu\dag, Alexandru Iosup\dag}
% \email{{s.battaglini-fischer, n.srinivasan, b.l.szarvas}@student.vu.nl, {x.chu, a.iosup}@vu.nl}
% \affiliation{%
%   \institution{Vrije Universiteit, Amsterdam, the Netherlands}
%   % \country{Amsterdam, the Netherlands}
% }

\author{S\'andor Battaglini-Fischer$^{*}$}
% \authornote{Authors contributed equally to this research.}
\affiliation{%
  \institution{Vrije Universiteit Amsterdam}
  \city{Amsterdam}
  \country{The Netherlands}
}
\email{s.battaglini-fischer@student.vu.nl}

\author{Nishanthi Srinivasan$^{*}$}
\affiliation{%
  \institution{Vrije Universiteit Amsterdam}
  \city{Amsterdam}
  \country{The Netherlands}
}
\email{n.srinivasan@student.vu.nl}

\author{B\'alint L\'aszl\'o Szarvas$^{*}$}
\affiliation{%
  \institution{Vrije Universiteit Amsterdam}
  \city{Amsterdam}
  \country{The Netherlands}
}
\email{b.l.szarvas@student.vu.nl}

\author{Xiaoyu Chu\dag}
\affiliation{%
  \institution{Vrije Universiteit Amsterdam}
  \city{Amsterdam}
  \country{The Netherlands}
}
\email{x.chu@vu.nl}

\author{Alexandru Iosup\dag}
\affiliation{%
  \institution{Vrije Universiteit Amsterdam}
  \city{Amsterdam}
  \country{The Netherlands}
}
\email{a.iosup@vu.nl}

\thanks{$^{*}$These authors contributed equally to this research.

\dag Corresponding authors.
}% <-this % stops a space

% \affiliation{%

% \affiliation{%

\renewcommand{\shortauthors}{S. Battaglini-Fischer, N. Srinivasan, B.L. Szarvas et al.}
\renewcommand{\shorttitle}{FAILS: Framework for Analysis of Incidents on LLM Services}

% \input{sections/0_Abstract}
\begin{abstract}
Large Language Model (LLM) services such as ChatGPT, DALL·E, and Cursor have quickly become essential for society, businesses, and individuals, empowering applications such as chatbots, image generation, and code assistance.
The complexity of LLM systems makes them prone to failures and affects their reliability and availability, yet their failure patterns are not fully understood, making it an emerging problem.
However, there are limited datasets and studies in this area, particularly lacking an open-access tool for analyzing LLM service failures based on incident reports.
Addressing these problems, in this work we propose \textit{FAILS}, the first open-sourced framework for incident reports collection and analysis on different LLM services and providers. 
\textit{FAILS} provides comprehensive data collection, analysis, and visualization capabilities, including:
(1) It can automatically collect, clean, and update incident data through its data scraper and processing components;
(2) It provides 17 types of failure analysis, allowing users to explore temporal trends of incidents, analyze service reliability metrics, such as Mean Time to Recovery (MTTR) and Mean Time Between Failures (MTBF);
(3) It leverages advanced LLM tools to assist in data analysis and interpretation, enabling users to gain observations and insights efficiently.
All functions are integrated in the backend, allowing users to easily access them through a web-based frontend interface.
\textit{FAILS} supports researchers, engineers, and general users to understand failure patterns and further mitigate operational incidents and outages in LLM services. 
The framework is publicly available on \url{https://github.com/atlarge-research/FAILS}.

% The code for this project is available on \href{https://github.com/sandor-battaglini-fischer/distributed-systems-lab/}{Github}.
% The framework is publicly available on \url{https://github.com/atlarge-research/FAILS}.
\end{abstract}





\maketitle

\begin{figure}[t]
    \centering
    \includegraphics[trim={0 72 0 72}, clip, width=\linewidth]{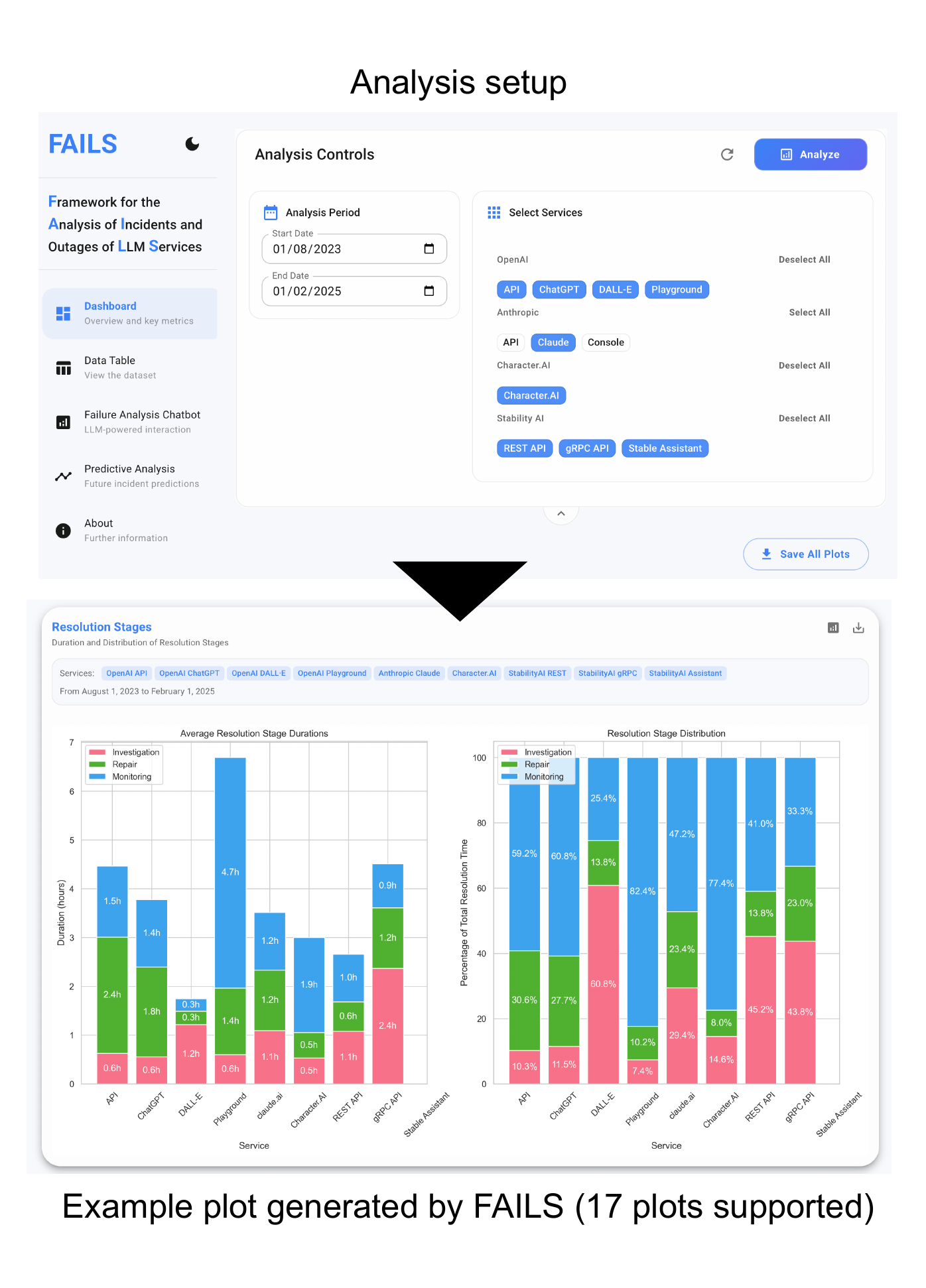}
    %\vspace*{-1cm}
    \caption{The frontend dashboard of \textit{FAILS} and one of the 17 resulting plots generated from it. 
    % In this figure, we depict FAILS capabilities in analyzing service incidents of 9 service providers. For readability and demonstration purposes, we present one plot generated by FAILS; for the complete list of plots supported by FAILS see \Cref{tab:plot_overview}.
    }
    \label{fig:intro:fails-interface}
    \vspace*{-0.8cm}
\end{figure}

\section{Introduction}\label{sec:introduction}

Large Language Models (LLMs) are transforming industries and societies through a broad range of applications, such as customer service \cite{Meduri2024Oct}, content generation \cite{Zhou2024Jan}, decision-making processes \cite{Xu2024Dec}, and scientific research \cite{Arman2023Sep}. 
LLM services operate as highly complex distributed systems, involving geographically dispersed components for training, inference, and user interaction \cite{chkirbene2024large}, which makes them prone to failures in infrastructure, software, and other external dependencies \cite{Yu2024Jun}. 
Failures and outages have major costs for users relying on LLM services, thus can cause significant degradation in customer loyalty, leading even the largest providers to issue a public apology after a system-wide outage \cite{observability_report, x_postopenai}. 
Therefore, understanding and mitigating these failures is a critical challenge to maintain the reliability and trustworthiness of these systems~\cite{Yu2024Jun}.

% \subsection{Motivation} 

Currently, tools for analyzing failures in LLM services are either private or have limited functionality. 
Companies such as SmartBear AlertSite~\cite{SmartbearAlertSite}, UptimeRobot~\cite{UptimeRobot}, and Pingdom~\cite{Pingdom} offer enterprise monitoring services, but are closed-source and aimed at business customers and system administrators. 
Downdetector~\cite{Downdetector} offers real-time problem and outage monitoring of public websites, but as a user-driven site lacks comparative tools and in-depth root-cause analysis, as well as the ability to select and compare different services from one provider. 
Although some work has been done to collect and investigate failures on these pages \cite{chu2025-llm-analysis, 9659508}, there is still a lack of tools that integrate data scraping, analysis, and visualization in an open-source framework for researchers, business owners, and the general public.

% LLM services are complex distributed systems designed to process and generate text, leveraging vast amounts of data and computational resources in the process. From a distributed systems perspective, LLM services consist of multiple interconnected components, including model storage, training pipelines, inference engines, data handling subsystems, and interfaces allowing users to interact with them. All of these components operate on geographically distributed machines and must work together to support the immense computing and memory requirements of training and serving models \cite{Brakel2024ModelPO, Duan2024EfficientTO}.
% This complexity and size come with significant challenges, including ensuring scalability as demand and models grow, managing energy consumption, and handling failures in their distributed components. These challenges are further compounded by external dependencies such as third-party APIs, load balancers, and content delivery networks. Analyzing service failures is a complex task, as self-reported incidents are often limited by company policies, and even companies themselves may struggle to identify the nature or origin of a failure immediately. Therefore, a tool for analyzing failures must be able to deal with dynamic data structures, offer granular insights, and be accessible.

% This leaves the self-reported status pages of the individual providers. 
% While publicly available, they lack standardization and analysis tools, complicating efforts to perform comparative analyses between different providers and services. 

% \subsection{Our Solution}

To address this challenge, we proposed \textbf{\textit{FAILS}} \textit{(\uline{\textbf{F}}ramework for \uline{\textbf{A}}nalysis of \uline{\textbf{I}}ncidents on \uline{\textbf{L}}LM \uline{\textbf{S}}ervices)}, which we designed and implemented to:
(1) Automatically scrape, clean, and store incident data from LLMs self-disclosed reports;
(2) Automatically perform a wide range of failure characterizations, such as distribution of MTTR and MTBF;
(3) Visualize data and analysis results, such as temporal failure trends, failure correlations, and recovery patterns;
(4) Integrate LLM tools to provide interactions with datasets, and contextual insights into analysis results;
(5) Provide a user-friendly and web-based interface to access all functions of the framework (See \Cref{fig:intro:fails-interface}).

% \begin{enumerate}
%     \item Automatically scrape, process, clean and store failure data from LLM provider status pages;
%     \item Visualize temporal failure trends, correlations and recovery patterns;
%     \item Perform statistical evaluations of reliability metrics, such as Mean Time to Recovery (MTTR) and Mean Time Between Failures (MTBF);
%     \item Integrate LLM-assisted methods to provide contextual insights into analysis results;
%     \item Present the user with an easy-to-use web-based interface to access all tools
% \end{enumerate}

Our key contributions are as follows:
\begin{enumerate}
    \item[\textbf{C1.}] \textbf{(Framework Design)} We design \textit{FAILS}, the first tool to automatically collect, monitor, and analyze incident reports from popular LLM providers and services. We describe the architecture and components of the framework, followed by the functional and non-functional requirements we formulated.
    \item[\textbf{C2.}] \textbf{(Implementation)} We implement \textit{FAILS}, which enables users to select and compare operational characteristics of different services and providers. We provide 17 types of failure analysis that can generate visual plots automatically. We also integrate LLM tools into the framework to allow users to interact with datasets and analyze plots directly in natural languages.
    \item[\textbf{C3.}] \textbf{(Open Science)} We follow the principles of open science and release the framework on \url{https://github.com/atlarge-research/FAILS}.
\end{enumerate}

% The remainder of this paper provides a brief background on LLMs, their issues and the requirements for an effective analysis tool. We then describe the architecture of the FAILS system, highlighting its core functionalities and implementation, and present a few selected experimental results obtained using it. Finally, we discuss potential future improvements and additional functionalities that could be added to this open-source project.

\section{Background}
In this section we provide an overview of LLM services, their failure recovery model, and the key concepts and metrics used in this work.

\subsection{LLM Services and Incidents}\label{sec:llm-service}
% \textcolor{red}{Explain what is LLM service, what types, and what we select for this work.}

We center our work on 11 LLM services from 4 providers, namely: (1) OpenAI (ChatGPT, API, DALL-E, and Playground), (2) Anthropic (Claude, API, and Console), (3) Character.AI, and (4) Stability.AI (RestAPI, gRPC API, and Stable Assistant). The services are selected because they have publicly accessible incident reports. 
% Once an incident occurs, a textual report of its failure recovery process is produced and disclosed to the public. 
These incident reports are valuable to understand the LLM availability and their failure recovery patterns.
% \subsection{The Complexity of LLM Services (MOVE TO \Cref{sec:introduction})}
% LLM services are complex distributed systems designed to process and generate text, leveraging vast amounts of data and computational resources in the process. From a distributed systems perspective, LLM services consist of multiple interconnected components, including model storage, training pipelines, inference engines, data handling subsystems, and interfaces allowing users to interact with them. All of these components operate on geographically distributed machines and must work together to support the immense computing and memory requirements of training and serving models \cite{Brakel2024ModelPO, Duan2024EfficientTO}.

% This complexity and size come with significant challenges, including ensuring scalability as demand and models grow, managing energy consumption, and handling failures in their distributed components. These challenges are further compounded by external dependencies such as third-party APIs, load balancers, and content delivery networks. Analyzing service failures is a complex task, as self-reported incidents are often limited by company policies, and even companies themselves may struggle to identify the nature or origin of a failure immediately. Therefore, a tool for analyzing failures must be able to deal with dynamic data structures, offer granular insights, and be accessible.

\subsection{Failure Recovery Model and Metrics}\label{sec:theoback}\label{sec:f-r-model}
\begin{figure}[t]
    \centering
    % \vspace*{-0.3cm}
    \includegraphics[width=\linewidth]{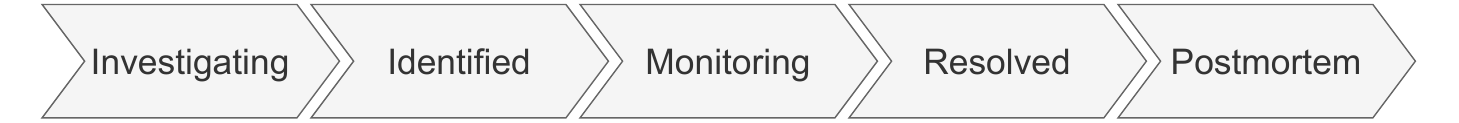}
        \caption{\textbf{Overview of the failure recovery model~\cite{chu2025-llm-analysis}.} 
        % The model outlines the 5 key statuses in order: Investigating, Identified, Monitoring, Resolved, and Postmortem.
        }
    \label{fig:frm}
      \vspace*{-0.6cm}
\end{figure}

The failure recovery analysis of incidents is grounded in the failure recovery model summarized in \cite{chu2025-llm-analysis}. 
This model provides a systematic framework to analyze how providers handle incidents, providing insight into fault tolerance and resilience.
% \cite{gamell2014exploring, avizienis2004basic}.
The adopted failure recovery model comprises five key statuses that describe the incident handling lifecycle, which is visualized in~\Cref{fig:frm}:
\begin{enumerate}[label=\textbf{S\arabic*}]
    \item \textbf{Investigating:} The operational team begins to analyze the issue immediately after it is observed.
    \item \textbf{Identified:} The problem of the incident is identified.
    \item \textbf{Monitoring:} A fix is implemented and the system is monitoring to ensure the resolution is effective.
    \item \textbf{Resolved:} The incident is considered resolved and normal operations resume.
    \item \textbf{Postmortem:} A detailed summary of the incident is presented to analyze the root cause and suggest measures to prevent recurrence.
\end{enumerate}

This model enables the following key metrics:

\textbf{MTBF (Mean Time Between Failures):} This metric calculates the average time between two consecutive failures for a provider. It serves as an indicator of how fault-tolerant the provider system is.

\textbf{MTTR (Mean Time To Recovery):} Derived from the failure recovery model, MTTR measures the total time taken to resolve a failure, encompassing the period from investigation (S1) to resolution (S4). Provides information on how quickly a provider can respond to and rectify failures.

\textbf{Co-occurrence of Failures:} This metric measures the number of services impacted simultaneously (service inter-dependencies) during a failure for each provider.
% It highlights service inter-dependencies and provides insights into the potential benefits of compartmentalization and dependency separation.

% \textcolor{red}{\textbf{Impact classification distribution:} Although the largest providers utilize a standardized framework to classify the severity of failures, variations exist in their classification tendencies. These differences reflect the operational priorities and reporting practices of the providers.}
\section{Design of \textit{FAILS}: a Tool for Automated Incident Data Collection and Analysis of LLM Services}
\label{sec:design}
In this section, we first analyze the requirements for \textit{FAILS}. Then, we present its architectural design, including its backend, frontend, and their interactions. Finally, we provide a detailed explanation of the implementation of different components and how they meet the proposed requirements.
% In order to address the requirements, we developed a system to facilitate the analysis of incidents and outages of Large Language Models services. Aptly named \textit{FAILS (Framework for Analysis of Incidents and Outages of LLM Services)}, it provides a web-based interface for automatically scraping the data from the providers and interacting with it in the form of tables, plots and text based outputs. Without loss of generality, we chose services from OpenAI (ChatGPT\cite{OpenAIChatGPT}, API service\cite{OpenAIAPI}, DALL-E\cite{OpenAIDALLE} and Playground\cite{OpenAIPlayground}), Anthropic (Claude\cite{AnthropicClaude}, API\cite{AnthropicAPI}, and Console\cite{AnthropicConsole}), CharacterAI\cite{CharacterAI} and StabilityAI (RestAPI\cite{StabilityRESTAPI}, gRPC API\cite{Stabilitygrpc} and Stable Assistent\cite{StabilityAssistent}) to demonstrate the viability and effectiveness of our solution.
\begin{figure*}[t]
  \centering
  \includegraphics[width=0.9\linewidth]{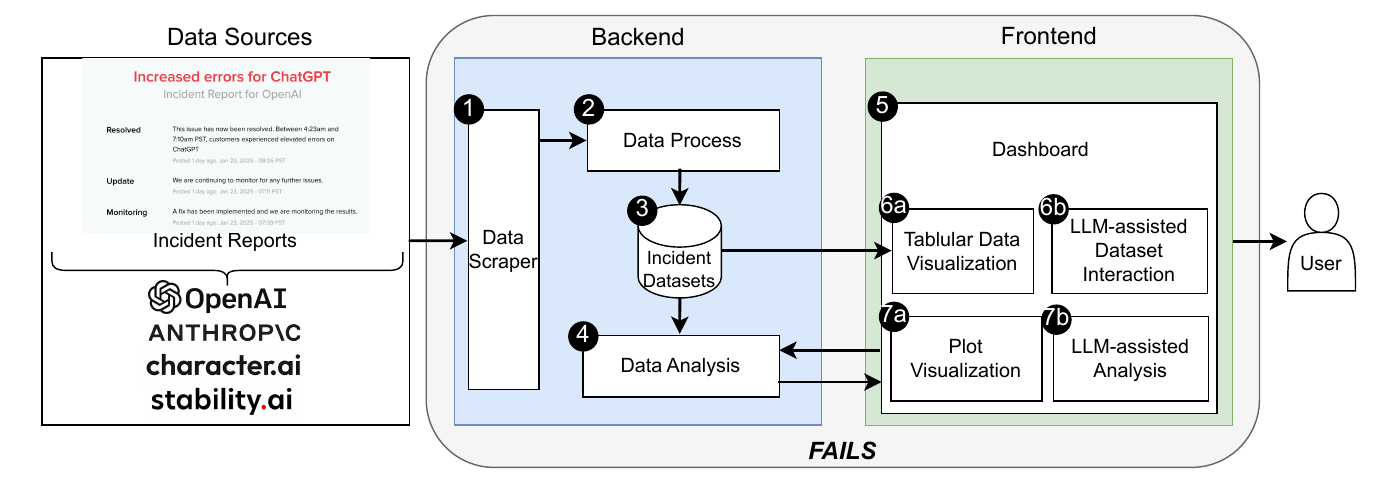}
  \vspace*{-0.3cm}
    \caption{\textbf{An overview of the architecture of \textit{FAILS}}.}
    \label{fig:architecture}
    \vspace*{-0.3cm}
\end{figure*}

\subsection{Requirement Analysis}\label{sec:design:requirement}\label{sec:requirement}
We have identified 5 functional requirements (\textbf{FR}), and 4 non-functional requirements (\textbf{NFR}) for \textit{FAILS}.

\begin{enumerate}
[label=\textbf{FR\arabic*}]
    \item \label{fr:1}\textbf{Data Collection}: The framework must be able to automatically scrape incident and outage data from multiple service providers. Given the distributed nature of these services, the pipeline must handle potential challenges such as incomplete or changing data structures. The framework should be able to automatically collect, process, and update the datasets according to the user configuration.  
    \item \label{fr:2}\textbf{Data Cleaning and Storage}: The collected data must be cleaned to ensure that the datasets are consistent and stored in a structured format. Data cleaning includes handling missing values, standardizing date-time formats, and ensuring compatibility between datasets from different providers. 
    % This step is essential to facilitate the downstream data analysis.
    \item \label{fr:3}\textbf{Failure Recovery Analysis}: The framework must be able to analyze the failure recovery process using common-used metrics such as MTTR and MTBF. Failure recovery analysis should allow users to identify failure recovery characteristics, such as evaluating how quickly services recover from failures, and compare failure recovery differences between multiple providers~\cite{gamell2014exploring, avizienis2004basic}.
    \item \label{fr:4}\textbf{Temporal and Co-occurrence Analysis}: The framework should support the exploration and comparison of failure trends over time, such as weekly failure distributions, and auto-relations. Additionally, the framework should support the identification of co-occurrence patterns between failures across multiple LLM services.
    % The framework should also support identifying the relationships between failures of different LLM services.
    \item \label{fr:5}\textbf{Data Visualization and Reporting}: The framework should present analytical results in a structured and interpretable manner. This includes generating graphical representations (e.g., time-series plots, heatmaps), tabular summaries, and textual reports that highlight key failure metrics. The focus of this requirement is on ensuring that the results of failure analysis are clearly structured and visually comprehensible for effective interpretation.
    %The datasets and results of the analysis must be presented in an accessible and unified format. This includes graphical visualizations (e.g., time-series plots, heatmaps), tabular summaries for structured data, and textual reports highlighting key failure metrics. Unlike \textbf{NFR1}, which focuses on the usability of the interface, this requirement is specifically concerned with the representation of analytical results in a meaningful and interpretable format.
    % This includes tables for structured data presentation, graphical visualizations, and textual summaries of key metrics.
\end{enumerate}

\begin{enumerate}
[label=\textbf{NFR\arabic*}]
    \item \label{nfr:1}\textbf{User-friendly and Accessible Web-based Interface}: The framework should provide an intuitive and interactive web-based interface that allows users to configure analyses, explore results, and compare failure patterns effortlessly. Unlike \textbf{FR5}, which pertains to how data is structured and represented, this requirement ensures that users, regardless of their technical expertise, can efficiently navigate and interact with the system. A modern, responsive design should support accessibility across different devices and platforms.
    %To ensure usability for diverse users, including nontechnical users, the framework should provide a modern, responsive, and intuitive web-based interface. This enables users to interact with advanced data processing and analysis without specialized knowledge and allows simple deployment with standard web technologies. Users should be able to select the service, provider, and types of analysis they want to perform and compare.
    \item \label{nfr:2}\textbf{Reliability and Fault Tolerance}: The framework should operate consistently without unexpected delays or errors. In particular, failures during data scraping and analysis could arise from new or inconsistent data formats from different LLM service providers. However, it is unfeasible to account for all errors, so implementing fault tolerance mechanisms, such as retry logic, data validation checks, and user-level error messages, is significant for the robustness and reliability of the system \cite{isukapalli2024systematic}.
    % \item \textbf{Performance and Scalability}: Ensuring the framework can run well on general web servers and consumer level machine is essential when it comes to effective distribution. Futureproofing the application by ensuring elasticity (vertical and horizontal scalability) is essential in the rapidly developing field of LLMs. Using generalized methodologies to scrape the data facilitates adding more services as they appear in the market, and making sure longer time-frames can be supported as well as adding value to the application.
    \item  \label{nfr:3}\textbf{Generality and Scalability:} To ensure the generality of our framework, we chose the following LLM providers and services:
    OpenAI (ChatGPT, API, DALL-E, and Playground), Anthropic (Claude, API, and Console), Character.AI, and Stability.AI (RestAPI, gRPC API, and Stable Assistant). FAILS can be easily scaled to other LLM providers and services if they use similar structured status pages. Additionally, it should be designed to handle growing data volumes and concurrent user access without performance degradation.
    \item \label{nfr:4}\textbf{LLMs-empowered Interaction:} Integrating the capabilities of LLM into the framework can provide more contextual insights from the failure patterns observed by the analysis results. LLM empowers users in root cause analysis and helps them understand more technical concepts. For example, the LLM can provide knowledge of the failure models used in the failure recovery analysis.
\end{enumerate}

\subsection{Design of \textit{FAILS} Architecture}
\textit{FAILS} follows a modern client-server architecture, consisting of front- and back-end. \Cref{fig:architecture} shows the architecture of the designed framework.
In the backend, \textit{FAILS} uses data scrapers (\bc{1}) to collect data on outages and incidents separately from the public status pages of OpenAI \cite{StatusOpenAI}, Anthropic \cite{StatusAnthropic}, Character.AI \cite{StatusCharacterAI}, Stability.AI \cite{StatusStabilityAI}. After a series of data cleaning and processing (\bc{2}), the raw data are stored as incident datasets (\bc{3}). The data analysis (\bc{4}) module provides various analysis functions that are called by user requests through the frontend dashboard (\bc{5}).

In the frontend, users can select the types of analysis to perform on selected service and provider through a dashboard (\bc{5}). Tabular data visualization (\bc{6a}) provides users with an overview of historical incidents, including information on the provider, service, impact, start and end time, and status. To support a better understanding of incident datasets, \textit{FAILS} integrates LLM-assisted dataset interaction (\bc{6b}), which allows users to inquire about the statistics of the dataset through a chatbot. The results of the analysis are presented through various figures in the visualization of the plot (\bc{7a}), and enhanced by an LLM-assisted analysis module (\bc{7b}). This module combines the plot images with predefined instructions to create a prompt to call the LLM API, providing users with a summary and insights from the analysis results.
% The frontend is built in React.js \cite{ReactJS} with Material UI \cite{MaterialUI} components, which provide a modern, responsive and intuitive user interface. This enables non-technical users to access the advanced data processing and analysis pipeline effectively and developers to easily modify the code of \textit{FAILS}, since React is the most popular JavaScript framework \cite{ReactPop} and Material UI is built on Google's design language.

% The backend encapsulates the data scraping, processing and analysis functionality with a Flask-based server, which is lightweight, popular and scalable \cite{Flask}.

% \begin{figure}[h!]
%     \centering
%     \includegraphics[width=\linewidth]{figures/Sys_Archietecture.png}
%         \caption{\textbf{Architecture of \textit{FAILS}}: a structural overview of the application, consisting of scraping functionality, the tabular data display, plot visualization, and AI tools in the backend, which are all accessible from a React-based web interface. The scraping pipelines are shown separately and generate a final CSV, which process powers all other functionality.}
%     \label{fig:architecture}
% \end{figure}

\subsection{Implementation of \textit{FAILS} Components} \label{sec:implementation}\label{sec:design:implement}

The implementation of the following features addresses the requirements in \Cref{sec:requirement}:

\subsubsection{Data scraping and processing pipeline}
\label{scraping}
The data scraping pipeline uses headless Chrome browsers through \textit{Selenium WebDriver} \cite{Selenium} to systematically extract historical outage and incident data from the status pages provided by service providers. It is widely supported, handles complex authorization well, and enables dynamic data retrieval by simulating user interactions, such as clicks and scrolls. The scraper utilizes \textit{WebDriverWait} conditions to ensure proper dynamic content loading and implements retry mechanisms to handle potential stale elements or network problems.
    
First, attributes and metadata such as title, impact level, and a color that represents severity are extracted from the historical incident list, along with an \textit{ID} to ensure uniqueness. The start and end time are also scraped for the entire failure and, where available, for the individual failure recovery stages (see Section \ref{sec:theoback}). 
% OpenAI, Anthropic, and Character.AI use status pages based on \textit{Atlassian Statuspage} \cite{Atlassian2025Jan} and can be handled within one pipeline. StabilityAI's page is handeled separately, due to being based on \textit{Instatus} \cite{Instatus}. 

Once extracted, the data undergoes a transformation process, where timestamps are normalized to \textit{UTC}. Service identification is handled through a parsing mechanism that recognizes both explicit service mentions, and implicit references in the incident descriptions. The system then processes and unifies the incident updates, which are stored as \textit{JSON} strings, to categorize the different stages of the incident. The processed dataset is stored in a separate \textit{CSV} file for each pipeline, and then merged after integrity and consistency checks. To prevent data corruption, file backups, temporary files and restore operations are also carried out automatically at this stage. The approach ensures that the pipeline achieves both reliability and scalability, addressing \ref{fr:1}, \ref{fr:2}, \ref{nfr:2}, and \ref{nfr:3}.

% As for Stability.AI, the scraper is designed for status pages from \textit{Instatus} \cite{Instatus}, and is structured differently. Aside from the page structure and location of metadata, Stability.AI has a unique set of services (REST API, gRPC API, Stable Assistant) that require specific detection logic. While the core scraping process is similar, we needed to ensure that the impact levels, colors, and failure messages are consistent, and do this through a careful translation process. The processed dataset is also saved to a temporary \textit{CSV} file.
    
% Before merging the data, the system performs thorough validation checks on both raw and transformed data, ensuring data integrity and structural consistency. The final processed data is saved to a combined CSV, with careful handling of file backups, temporary files and restore operations to prevent data corruption. The system maintains separate archives for raw data from each provider while combining them into a unified, normalized format in the final output file. This standardized file supports analysis and visualization, aligning records across services based on timestamps and incident IDs. The approach ensures the pipeline achieves both reliability and scalability, addressing \ref{fr:1}, \ref{fr:2}, \ref{nfr:2}, and \ref{nfr:3}.

% The entire process can be launched from a single script, which also handles errors and ensures unique data entries. Thus, integration in the front end is simplified to a single API endpoint that triggers the entire scraping process and handles errors and retries gracefully.

\begin{figure}[t]
  \centering
  \begin{minipage}{\textwidth}
  \begin{subfigure}[b]{0.5\textwidth}
    \includegraphics[width=0.9\linewidth]{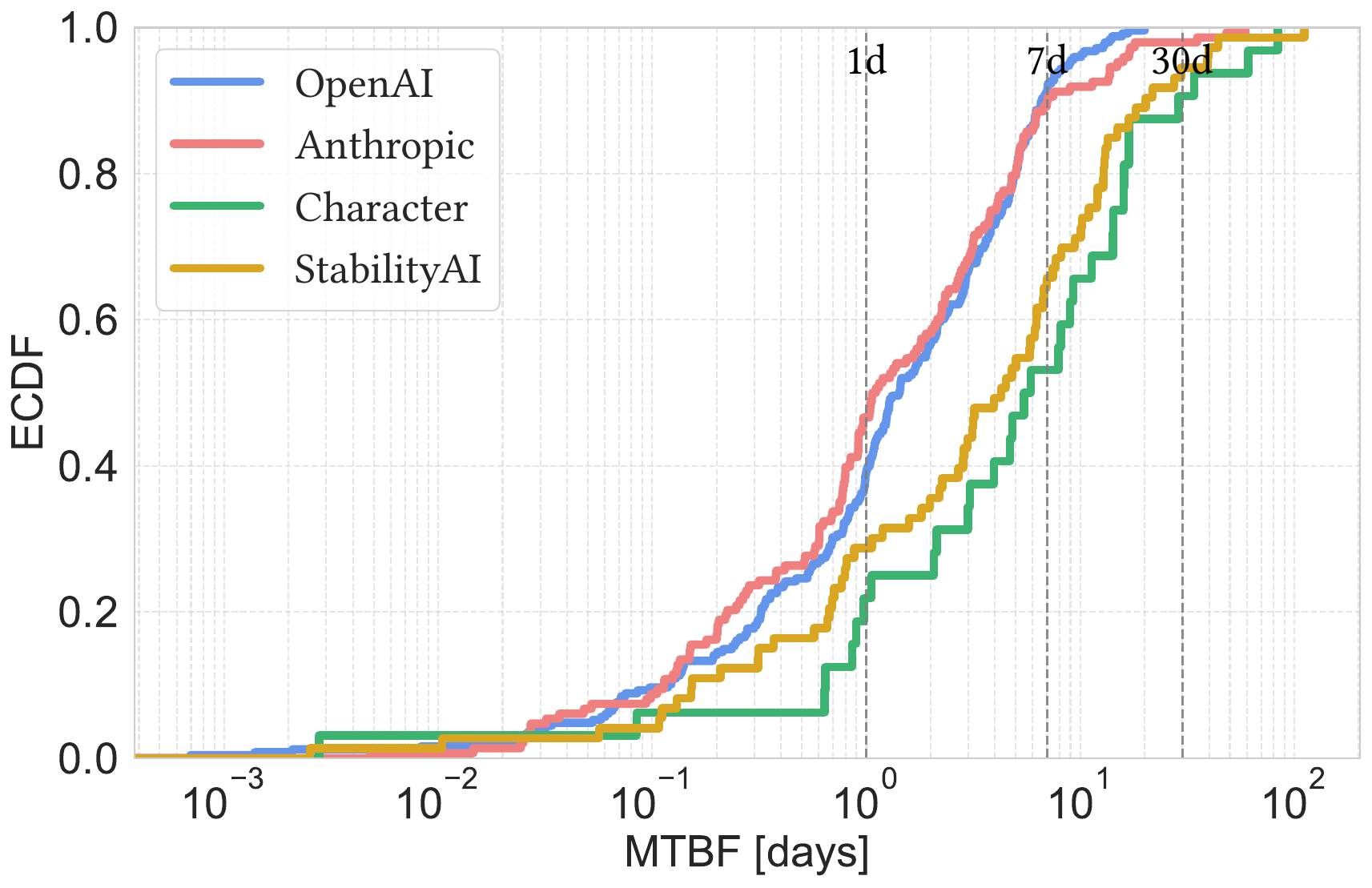}
    \vspace*{-0.2cm}
    \caption{Mean Time Between Failures (Longer is better).}
    \label{fig:mtbf}
  \end{subfigure}
  \vskip\baselineskip
  \vspace*{-0.3cm}
  \begin{subfigure}[b]{0.5\textwidth}
    \includegraphics[width=0.9\linewidth]{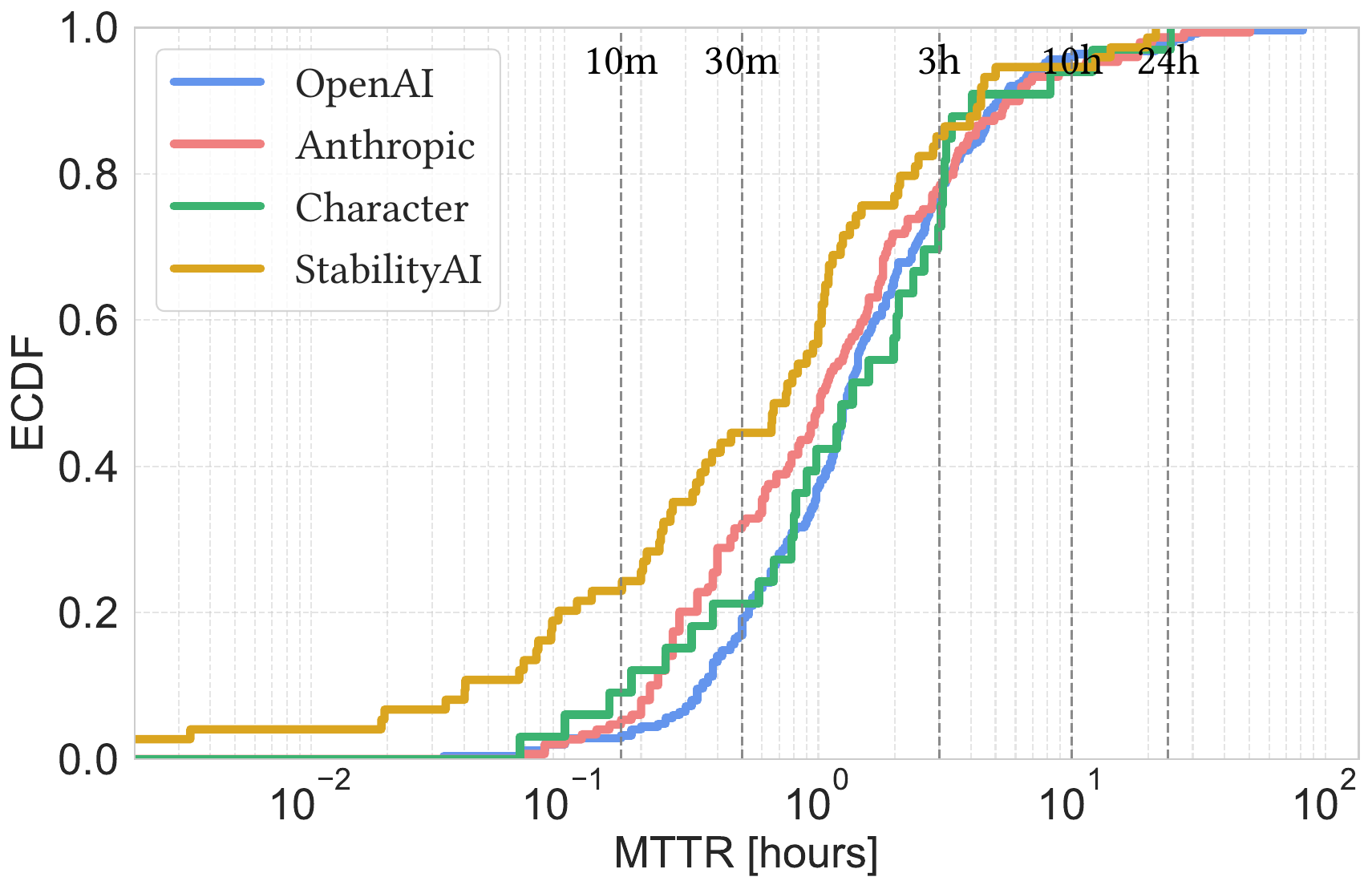}
    \vspace*{-0.2cm}
    \caption{Mean Time To Resolve (Shorter is better).}
    \label{fig:mttr}
  \end{subfigure}
  \end{minipage}
  \vspace{-0.3cm}
  \caption{CDF plot of MTBF and MTTR by provider.}
  \label{fig:mttr-and-mtbf}
  \vspace{-0.3cm}
\end{figure}
    
\subsubsection{Tabular data visualization}
The CSV format in which the data are saved enables the effective display using a table, such as in \textit{Material-UI's DataGrid} \cite{MUIDataGrid}. We used this to display select properties so that users can easily search for information on the latest recorded failures. Through color classifications, users can quickly identify services and impact levels, and read the precise error messages as reported by the providers, all on one page. Sorting by dates, durations, providers, and impact levels helps to provide an overview. It also helps investigate and verify the results of the analysis. 
In addition, the data scraping pipeline can be started from a button, which automatically updates the table at a fixed time interval. This addresses \ref{fr:5}.

\subsubsection{Automated plotting}
Our "Dashboard" page has the functionality to select a time frame, and a set of individual services from the dataset and plot metrics from this subset of data in an automated layout (addresses \ref{fr:3}, \ref{fr:4}, and \ref{fr:5}). The generated plots can be easily downloaded individually or in bulk as high-resolution PNGs, ready for sharing or use in documents and presentations. The complete list of the plots auto-producted by \textit{FAILS} can be found in \Cref{sec:plot-list}, \Cref{tab:plot_overview}.

\subsubsection{LLM-assisted plot analysis}

To aid in the interpretation of the generated plots, especially for a nonscientific audience, we implemented an automatic analysis pipeline, which addresses \ref{nfr:4}. It enables users to generate an automatic analysis of the current individual plots. This works by sending the image to an LLM service (in our test case, the API of OpenAI's \textit{gpt-4o-mini}), packed in a prompt informed by the knowledge of the individual graph. For example:

% \begin{tcolorbox}[enhanced, breakable]
\begin{quote}
\textit{Analyze this impact level distribution, with a focus on differences in distributions between services.
The distribution information: \\
\textbf{\textsc{\{date\_range, selected\_services, image\}}}.  \\
The impact levels are defined as \textbf{\textsc{\{impact\_definition\}}}, 
The calculated statistical metrics are: \\ \textbf{\textsc{\{statistical\_measures (mean, median, number of incidents)\}}}.}
\end{quote}
% \end{tcolorbox}

Additional prompts ensure a structured output of a predefined length, which enables us to format and show the response in the front-end. Additionally, we implemented an option to analyze all the plots in bulk, by passing the entire set of generated plots with a prompt, telling the model to focus on the most significant findings.

\subsubsection{Interacting with the dataset using LLMs}
% The aforementioned automated plot analysis has its limitations as it relies on predefined prompts and pre-calculated metrics. 
To add a more dynamic way of interacting with the data (\ref{nfr:4}), we input a cleaned version of the dataset and its metrics into the LLM API. In this way, users can interact with a chatbot interface to ask targeted questions about the dataset, such as the number of incidents, the time range of the dataset, and the root causes of individual events. Especially in conjunction with the data table and the generated plots, this enables deep insight into the data. Further experimentation with different prompting paradigms could result in a more informative output from the LLM \cite{white2023prompt}.
    
\subsubsection{Functionality wrapped in a web interface}
All of this functionality is neatly wrapped in a web-app interface (\ref{nfr:1}), which includes modern functionality such as theming and responsiveness. The individual functions mentioned above are arranged in a sidebar menu, which is clean and easily accessible and includes a page with information on the project.

% \subsection{Design process}
% The design process followed a structured approach to ensure clarity, collaboration, and effective execution. The first step involved maintaining a comprehensive to-do list to track progress and prioritize tasks, which were discussed with our TA in weekly meetings. We began by analysing the problem from a high-level perspective, then setting out the functional and non-functional requirements based on the problem and user needs. We checked the market for similar tools too, such as Grafana\cite{Grafana} for monitoring Distributed Systems, Datadog\cite{Datadog} for incident tracking, and Splunk\cite{Splunk} for log analysis, as well as the tools mentioned in the introduction. These examples provided inspiration and highlighted gaps in existing solutions that our system could address. A working design document was created to outline the system architecture, core components, and implementation strategy, which served as a reference throughout the development process.

% Tasks were divided based on team members' areas of expertise to maximize efficiency. Regular meetings and iterative feedback loops ensured that we remained on track, allowing adjustments to the design as ideas emerged. Thus, we were able to successfully translate the initial requirements into a robust and functional system.
\section{Experimental Results}\label{sec:experiment}
In this section, we select 3 showcases to demonstrate \textit{FAILS}'s functions, including: (1) The collected incident dataset by \textit{FAILS}; (2) We give 2 types of analysis as examples: MTBF and MTTR distributions, and co-occurrence of failures; (3) The dataset interaction through a LLM-enabled chatbot.

\subsection{LLM Incidents Dataset Collected by \textit{FAILS}} \label{sec:dataset}
To prepare the dataset used for failure analysis, we scraped and processed the data from the selected providers' status pages using \textit{FAILS}, according to the process described in section \Cref{scraping}. The resulting dataset includes all history incident reports provided by the LLM providers, starting from the date the first incident was reported until the day of scraping (2025-01-10 13:03:25). 

The collected LLM incidents dataset is summarized in \Cref{tab:provider_reports}. The incident reports span a time range from approximately 11 months (Stability.AI) to nearly 4 years (OpenAI), based on the release dates of their services.

\begin{figure*}[t]
    \centering
    \vspace*{-0.3cm}
    \includegraphics[width=0.3\textwidth]{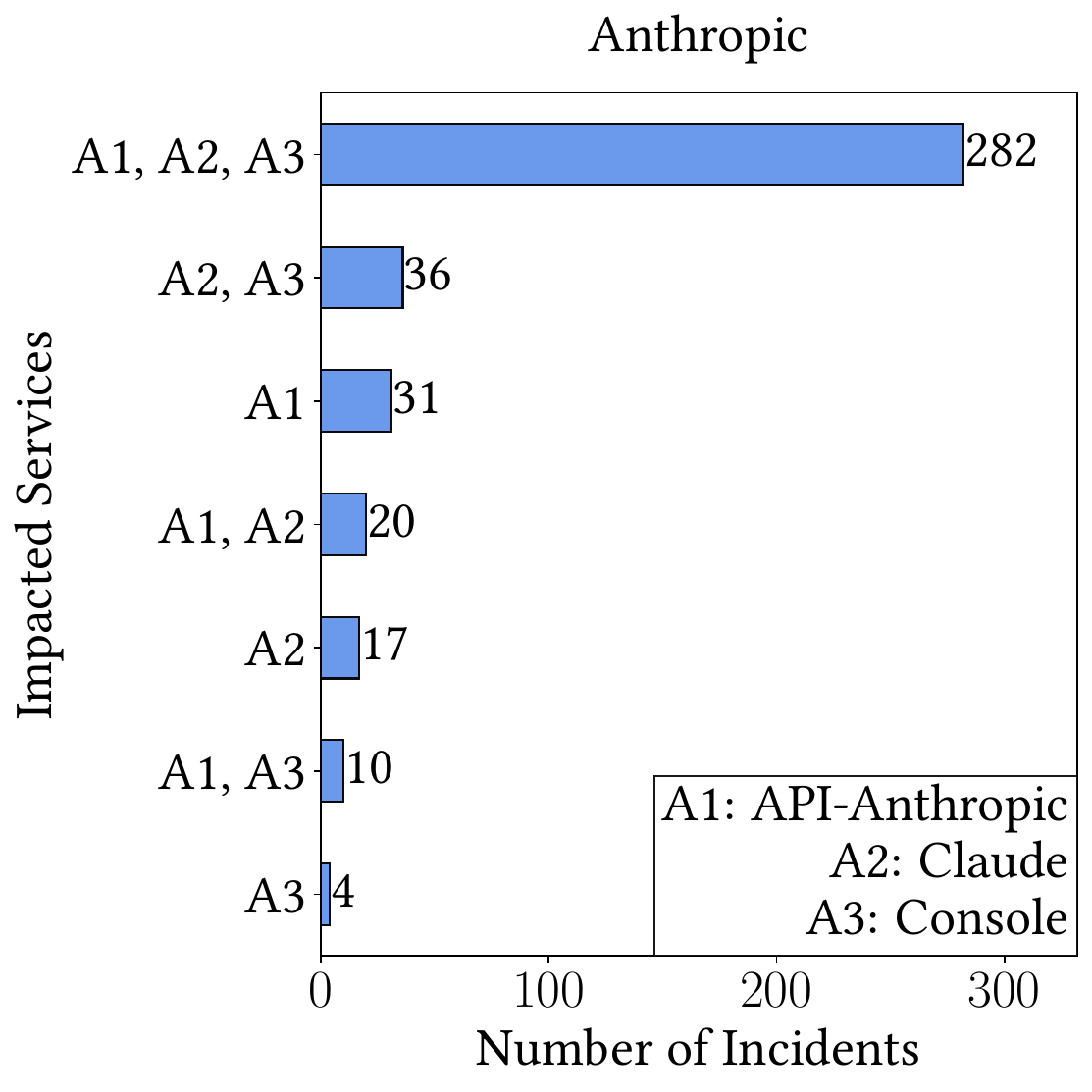}
    \includegraphics[width=0.3\textwidth]{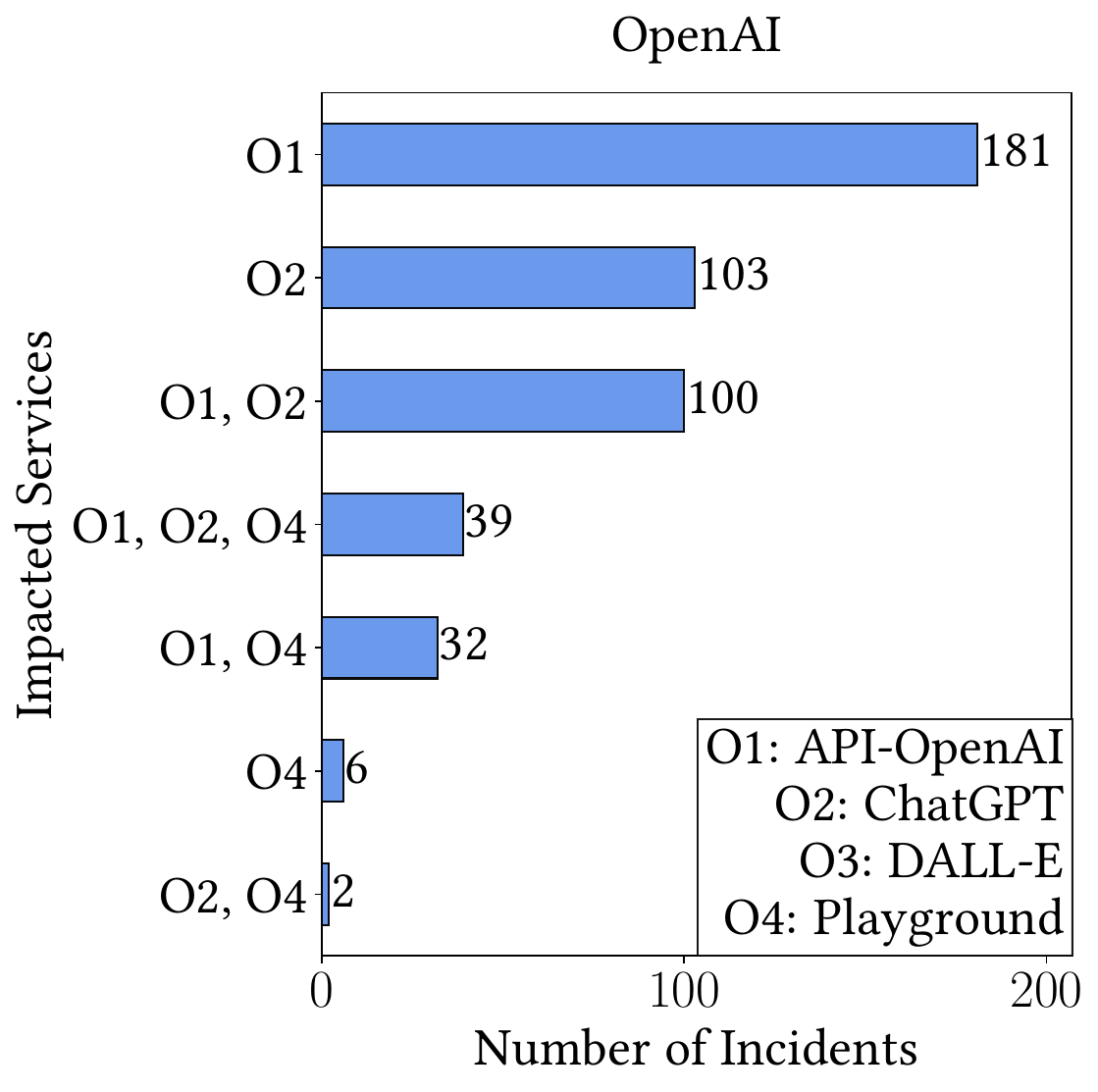}
    \includegraphics[width=0.3\textwidth]{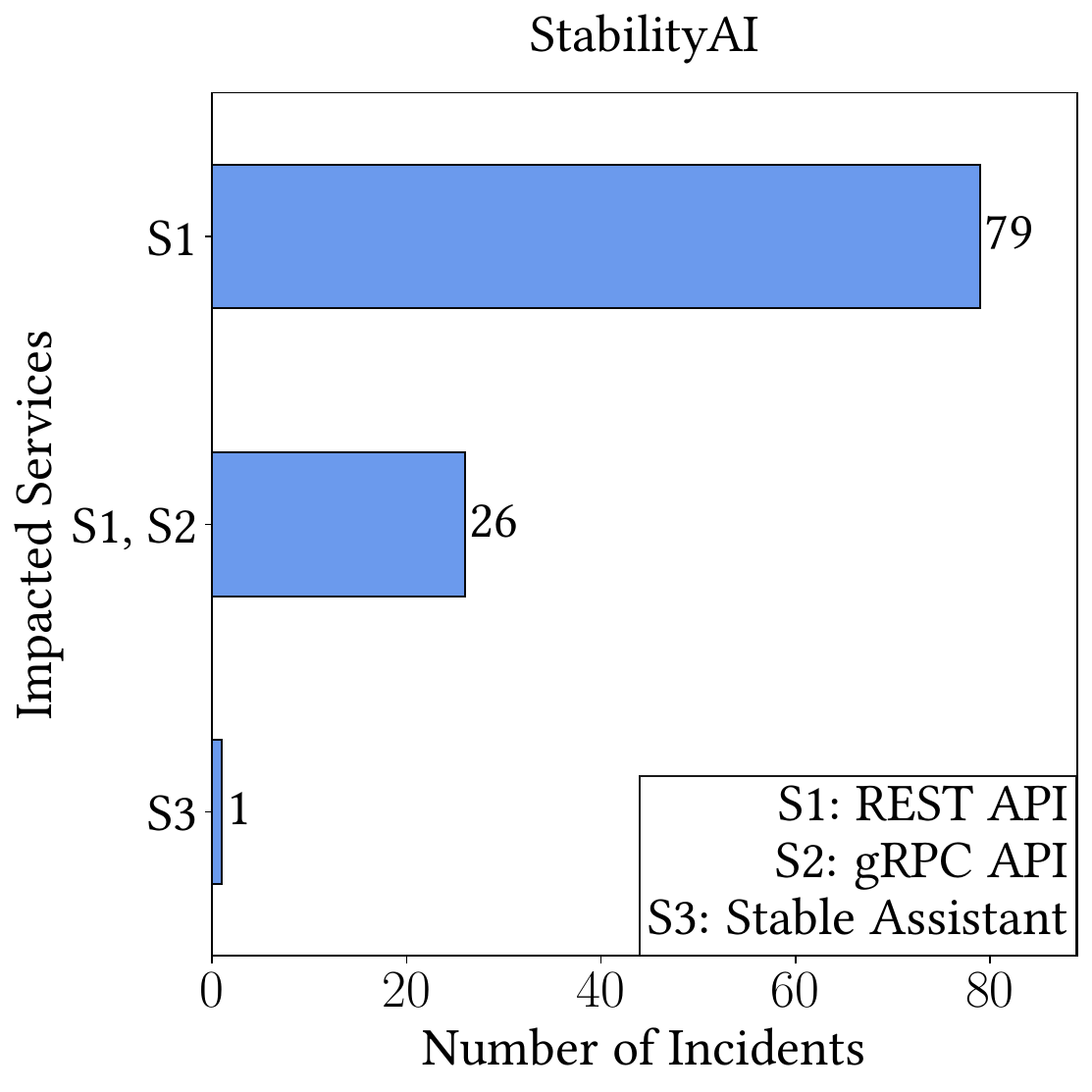}
    \vspace*{-0.3cm}
    \caption{\textbf{Co-occurrence of failures across services for each provider.} 
    }
    \label{fig:co-occurence}
    \vspace*{-0.3cm}
\end{figure*}

\begin{table}[t]
\caption{Summary of the LLM incident dataset collected on 2025-01-10.
% , including the first and last recorded dates and the total number of reports for each provider.
}
\vspace*{-0.3cm}
\centering
\renewcommand{\arraystretch}{1.5}
\small
\begin{tabular}{lccc}
\toprule
\textbf{Provider} & \textbf{First Date} & \textbf{Last Date} & \textbf{Number of Reports} \\
\midrule
OpenAI        & 2021-02-09  & 2025-01-08  & 414 \\
Anthropic     & 2023-03-16  & 2025-01-09  & 225 \\
Stability.AI   & 2023-01-24  & 2024-12-19  & 94  \\
Character.AI     & 2023-10-24  & 2025-01-10  & 59  \\
\bottomrule
\end{tabular}
\vspace*{-0.3cm}
\label{tab:provider_reports}
\end{table}

% \vspace{-4mm}

\subsection{Failure Analysis of LLMs Service Incidents} \label{sec:analysis}
We present the following failure analysis results, which are generated automatically by \textit{FAILS}.

\subsubsection{MTBF and MTTR}
\Cref{fig:mtbf} displays the Cumulative Distribution Function (CDF) of the MTBF in days for each provider. A steeper CDF indicates that MTBF values are less evenly distributed, while a distribution function shifted further to the right signifies longer times between failures for the provider.

Notably, Character.AI and Stability.AI exhibit longer MTBFs and less steep CDFs compared to OpenAI and Anthropic, which is likely due to the higher complexity and user demand faced by the latter two providers. 
In contrast, smaller user bases and simpler system designs likely contribute to the extended MTBFs of Character.AI and Stability.AI.

% \begin{figure}[H]
%     \centering
%     \includegraphics[trim=30 9 2 8, clip, width=1\linewidth]{figures/mtbf_provider.pdf}
%     \caption{\textbf{Cumulative Distribution Function (CDF) of the Mean Time Between Failures (MTBF) for Each Provider from February 2021 to December 2024} A steeper CDF indicates that MTBF values are less evenly distributed, while a distribution function shifted further to the right signifies longer times between failures for the provider. Notably, CharacterAI and StabilityAI exhibit longer MTBFs and less steep CDFs compared to OpenAI and Anthropic.}
%     \label{fig:mtbf}
% \end{figure}

% \subsubsection{Mean Time to Recovery (MTTR)}
The CDF distributions of the MTTR in hours for each provider are presented in \Cref{fig:mttr}. A steeper CDF indicates less evenly distributed MTTR values, while a distribution function shifted further to the right represents longer recovery times for the provider.
Stability.AI exhibits a left-shifted CDF, indicating significantly shorter recovery times than OpenAI, Anthropic, and Character.AI. 
% This outcome suggests that a smaller operational scale may facilitate better fault-tolerant systems.

% \begin{figure}[H]
%     \centering
%     \includegraphics[trim=30 9 2 8, clip, width=1\linewidth]{figures/mttr_provider.pdf}
%     \caption{\textbf{Cumulative Distribution Function (CDF) of the Mean Time to Recovery (MTTR) for Each Provider  from February 2021 to December 2024}. A steeper CDF indicates less evenly distributed MTTR values, while a distribution function shifted further to the right represents longer recovery times for the provider. StabilityAI exhibits a left-shifted CDF compared to the other three, indicating significantly shorter recovery times.}
%     \label{fig:mttr}
% \end{figure}

\subsubsection{Co-occurrence of Failures}
\Cref{fig:co-occurence} highlights the number of services affected simultaneously during failures for each provider. 
The subplots illustrate the number of services impacted during a failure: (a) For Anthropic, failures most frequently impact all three services simultaneously; 
(b) For OpenAI, failures tend to impact a single service more often than multiple services;
(c) For Stability.AI, failures predominantly affect only one service at a time.

In summary, Anthropic frequently experiences failures affecting multiple services, suggesting significant inter-dependencies among its systems. In contrast, OpenAI and Stability.AI tend to experience failures that impact only one service, indicating better compartmentalization and reduced cascading risks.

\subsection{Dataset Interaction through LLM Chatbot}
\begin{figure}[t]
  \centering
      \includegraphics[width=0.99\linewidth]{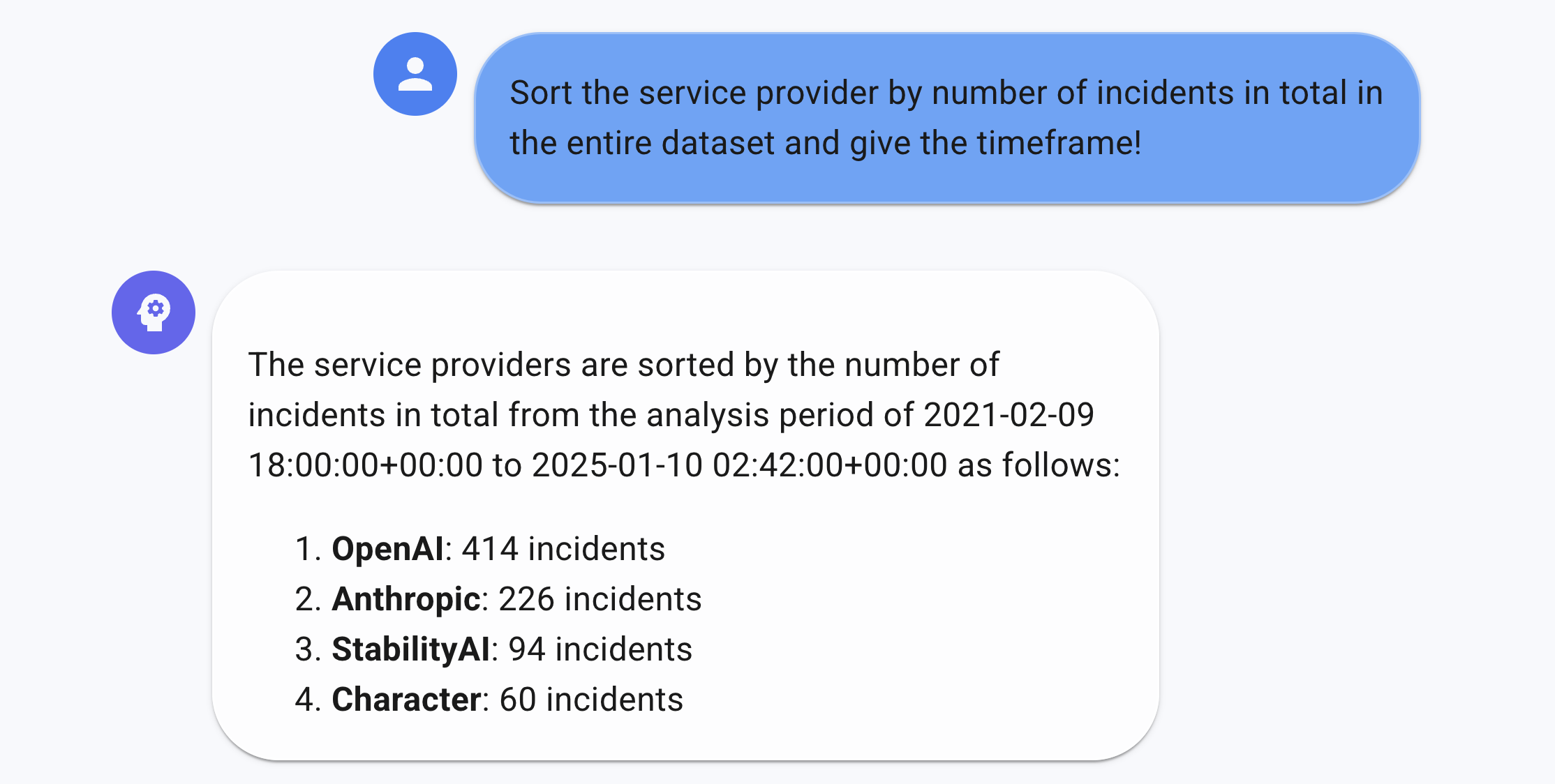}
  \vspace{-0.2cm}
        \caption{\textbf{The chatbot in \textit{FAILS}.} Here users can interact with the dataset in natural language.}
       \label{fig:chatbot}
    \vspace*{-0.3cm}
\end{figure}

The LLM-assisted analysis feature (\bc{6b} in \Cref{fig:architecture}), accessible from the sidebar of the frontend dashboard, enables users to query the dataset through a chatbot using natural language in a conversational format.
\Cref{fig:chatbot} shows an example use case, where the user asked the chatbot about the time range and total number of incidents.
The chatbot responds based on the provided dataset, answering the user's questions.

However, the accuracy and quality of the response depend on the chosen LLM, the prompts used, and are constrained by the LLM's limitations. This limitation of our work will be further discussed in \Cref{sec:limitations}.

\section{Limitations and Validity}\label{sec:limitations}
Our framework provides a foundation for effective analysis of LLM service failures.
However, our approach has certain limitations:

\textbf{The incompleteness of datasets: } Despite gathering incident data from four LLM providers and eleven services, the fact that these data are self-reported by operators poses risks of inaccuracy or incomplete disclosure. Some incidents may occur without being detected by the operational team. Moreover, companies may choose not to disclose all incidents due to potential impacts on market interests and policies.
% \textcolor{red}{[REF]}.

\textbf{The accuracy of LLM-assisted tools:} For the interaction and analysis of LLM-assisted datasets, the accuracy of LLM-generated content is not explicitly evaluated. As a result, the reliability of the output solely depends on the performance of the LLM APIs, raising concerns that it may not accurately reflect the characteristics of the data and plots.

\section{Related Work}\label{sec:related-work}

In the relatively new field of LLMs, previous work has begun to address the reliability of these public services by analyzing their outages and failure-recovery processes. For example, previous wrok \cite{chu2025-llm-analysis} conducted an empirical characterization of outages and incidents in public LLM services, providing valuable insights into metrics such as MTBF and MTTR as well as temporal patterns and co-occurrences. While this work sets the foundation for our work, it focuses primarily on establishing the metrics needed for the analysis without offering a practical, interactive, and extendable tool for further exploration or visualization of such data. 
% Additionally, our approach using LLMs themselves to aid in the analysis is to date unique in this field.

In the broader context of failure characterization, previous work has extensively explored failures across various related domains, such as HPC \cite{Park2023Jun, HPCfailures}, machine learning \cite{10763608, 2023-hotcloudperf-mlfailures}, cloud computing \cite{Cetin2021Oct}, and other large-scale distributed systems \cite{Yigitbasi2010Oct, Gupta2017Nov}.
While these studies have improved fault tolerance mechanisms, they are not directly tailored to the operational challenges and complexities of LLM services. In-domain studies such as the dynamic of the response and requests \cite{Wang2024Jan}, training \cite{Hu2024Characterization}, and inference \cite{Lazuka2024Oct} focus primarily on the performance and scalability aspects of LLM workloads rather than the characterization of failures and recovery processes.

Our framework fills the gap in the current corpus of work on failure analysis of LLM services by providing an open-source tool for researches that combines well-known failure characterization metrics with a modern, LLM assisted approach. 

% \textbf{Failure Characterization}

% \textbf{LLM Workload Characterization}

\section{Conclusion and Future Work}
\textit{FAILS} enhances failure analysis in complex LLM ecosystems by integrating automated data scraping, advanced analytics, and intuitive visualizations. With real-time data integration, predictive modeling, and geographic user-reported incident analysis, it helps improve reliability and decision-making in large-scale AI systems.

The \textit{FAILS} streamlines failure analysis by integrating data processing, visualization, and analytics. Its robust scraping pipeline ensures reliable, up-to-date data across providers, enabling detailed studies of MTBF, MTTR, and failure co-occurrence patterns.
The interactive dashboard allows users to explore failure trends dynamically. Automated CDF visualizations highlight recovery time variations, while LLM-assisted analysis provides deeper insights into provider-specific resilience. The intuitive design of the framework enables rapid visualization adjustments, allowing both technical and non-technical users to easily conduct meaningful comparisons.
By reducing manual effort and enhancing cross-provider comparisons, \textit{FAILS} provides actionable insights that contribute to improving system resilience and reliability.

Our future work includes: (1) Real-time prediction: Enhancing \textit{FAILS} with real-time failure tracking will open new possibilities, such as predicting recovery times upon the first report of downtime. By modeling provider, service, location, and historical data as a feature vector, a transformer-based approach can improve response strategies.
(2) RAG-enhanced analysis: Strengthening LLM-driven insights through refined prompts and a \textit{Retrieval Augmented Generation (RAG)} system will enhance contextual understanding. Integrating local models instead of APIs will improve fault tolerance, ensuring \textit{FAILS} remains fully functional even during service disruptions.
(3) Comparison with user reports: Extending the failure analysis with user-reported data from third-party platforms (e.g. Downdetector or X/Twitter) will provide a more comprehensive view of incidents. Advancing our approach to dynamic graph scraping will enable deeper comparisons between provider and user-reported failures, improving the data transparency and accuracy.

\newpage
% \begin{acks}
% TODO.
% \end{acks}

\bibliographystyle{ACM-Reference-Format}
\bibliography{reference}


\begin{thebibliography}{34}

%%% the Web address.

\ifx \showCODEN    \undefined \def \showCODEN     #1{\unskip}     \fi
\ifx \showISBNx    \undefined \def \showISBNx     #1{\unskip}     \fi
\ifx \showISBNxiii \undefined \def \showISBNxiii  #1{\unskip}     \fi
\ifx \showISSN     \undefined \def \showISSN      #1{\unskip}     \fi
\ifx \showLCCN     \undefined \def \showLCCN      #1{\unskip}     \fi
\ifx \shownote     \undefined \def \shownote      #1{#1}          \fi
\ifx \showarticletitle \undefined \def \showarticletitle #1{#1}   \fi
\ifx \showURL      \undefined \def \showURL       {\relax}        \fi
% invisible to TeX
\providecommand\bibfield[2]{#2}
\providecommand\bibinfo[2]{#2}
\providecommand\natexlab[1]{#1}
\providecommand\showeprint[2][]{arXiv:#2}

\bibitem[Sma(2018)]%
        {SmartbearAlertSite}
 \bibinfo{year}{2018}\natexlab{}.
\newblock \bibinfo{title}{{AlertSite Software as a Service - Everbridge at SmartBear Connect}}.
\newblock
\urldef\tempurl%
\url{https://smartbear.com/product/alertsite}
\showURL{%
\tempurl}
\newblock
\shownote{[Accessed 14. Jan. 2025]}.


\bibitem[Pin(2024)]%
        {Pingdom}
 \bibinfo{year}{2024}\natexlab{}.
\newblock \bibinfo{title}{{Pingdom}}.
\newblock
\urldef\tempurl%
\url{https://www.pingdom.com}
\showURL{%
\tempurl}
\newblock
\shownote{[Accessed 14. Jan. 2025]}.


\bibitem[Upt(2024)]%
        {UptimeRobot}
 \bibinfo{year}{2024}\natexlab{}.
\newblock \bibinfo{title}{{UptimeRobot: Free Website Monitoring Service}}.
\newblock
\urldef\tempurl%
\url{https://uptimerobot.com}
\showURL{%
\tempurl}
\newblock
\shownote{[Accessed 14. Jan. 2025]}.


\bibitem[Sel(2025)]%
        {Selenium}
 \bibinfo{year}{2025}\natexlab{}.
\newblock \bibinfo{title}{{About Selenium}}.
\newblock
\urldef\tempurl%
\url{https://www.selenium.dev/about}
\showURL{%
\tempurl}
\newblock
\shownote{[Accessed 14. Jan. 2025]}.


\bibitem[Sta(2025a)]%
        {StatusAnthropic}
 \bibinfo{year}{2025}\natexlab{a}.
\newblock \bibinfo{title}{{Anthropic Status}}.
\newblock
\urldef\tempurl%
\url{https://status.anthropic.com}
\showURL{%
\tempurl}
\newblock
\shownote{[Accessed 13. Jan. 2025]}.


\bibitem[Sta(2025b)]%
        {StatusCharacterAI}
 \bibinfo{year}{2025}\natexlab{b}.
\newblock \bibinfo{title}{{Character AI Status Status}}.
\newblock
\urldef\tempurl%
\url{https://status.character.ai}
\showURL{%
\tempurl}
\newblock
\shownote{[Accessed 13. Jan. 2025]}.


\bibitem[Dow(2025)]%
        {Downdetector}
 \bibinfo{year}{2025}\natexlab{}.
\newblock \bibinfo{title}{{Downdetector}}.
\newblock
\urldef\tempurl%
\url{https://downdetector.com}
\showURL{%
\tempurl}
\newblock
\shownote{[Accessed 14. Jan. 2025]}.


\bibitem[Sta(2025c)]%
        {StatusOpenAI}
 \bibinfo{year}{2025}\natexlab{c}.
\newblock \bibinfo{title}{{OpenAI Status}}.
\newblock
\urldef\tempurl%
\url{https://status.openai.com}
\showURL{%
\tempurl}
\newblock
\shownote{[Accessed 13. Jan. 2025]}.


\bibitem[MUI(2025)]%
        {MUIDataGrid}
 \bibinfo{year}{2025}\natexlab{}.
\newblock \bibinfo{title}{{React Data Grid component - MUI X}}.
\newblock
\urldef\tempurl%
\url{https://mui.com/x/react-data-grid/?srsltid=AfmBOoozaD7Vh9WjEftiJHA_6H7zcC6IxiDtbtqjVAlI_JDkBaDBdTsi}
\showURL{%
\tempurl}
\newblock
\shownote{[Accessed 13. Jan. 2025]}.


\bibitem[Sta(2025d)]%
        {StatusStabilityAI}
 \bibinfo{year}{2025}\natexlab{d}.
\newblock \bibinfo{title}{{Stability AI Platform - Status}}.
\newblock
\urldef\tempurl%
\url{https://stabilityai.instatus.com}
\showURL{%
\tempurl}
\newblock
\shownote{[Accessed 13. Jan. 2025]}.


\bibitem[Arman and Lamiyar(2023)]%
        {Arman2023Sep}
\bibfield{author}{\bibinfo{person}{Md Arman} {and} \bibinfo{person}{Umama Lamiyar}.} \bibinfo{year}{2023}\natexlab{}.
\newblock \showarticletitle{{Exploring the Implication of ChatGPT AI for Business: Efficiency and Challenges}}.
\newblock \bibinfo{journal}{\emph{International Journal of Marketing and Digital Creative}}  \bibinfo{volume}{1} (\bibinfo{date}{Sept.} \bibinfo{year}{2023}), \bibinfo{pages}{64--84}.
\newblock
\href{https://doi.org/10.31098/ijmadic.v1i2.1872}{doi:\nolinkurl{10.31098/ijmadic.v1i2.1872}}


\bibitem[Avizienis et~al\mbox{.}(2004)]%
        {avizienis2004basic}
\bibfield{author}{\bibinfo{person}{Algirdas Avizienis}, \bibinfo{person}{J-C Laprie}, \bibinfo{person}{Brian Randell}, {and} \bibinfo{person}{Carl Landwehr}.} \bibinfo{year}{2004}\natexlab{}.
\newblock \showarticletitle{Basic concepts and taxonomy of dependable and secure computing}.
\newblock \bibinfo{journal}{\emph{IEEE transactions on dependable and secure computing}} \bibinfo{volume}{1}, \bibinfo{number}{1} (\bibinfo{year}{2004}), \bibinfo{pages}{11--33}.
\newblock


\bibitem[Cetin et~al\mbox{.}(2021)]%
        {Cetin2021Oct}
\bibfield{author}{\bibinfo{person}{Mehmet~Berk Cetin}, \bibinfo{person}{Sacheendra Talluri}, {and} \bibinfo{person}{Alexandru Iosup}.} \bibinfo{year}{2021}\natexlab{}.
\newblock \showarticletitle{{Characterizing User and Provider Reported Cloud Failures}}.
\newblock \bibinfo{journal}{\emph{arXiv}} (\bibinfo{date}{Oct.} \bibinfo{year}{2021}).
\newblock
\href{https://doi.org/10.48550/arXiv.2110.12237}{doi:\nolinkurl{10.48550/arXiv.2110.12237}}
\showeprint{2110.12237}


\bibitem[Chkirbene et~al\mbox{.}(2024)]%
        {chkirbene2024large}
\bibfield{author}{\bibinfo{person}{Zina Chkirbene}, \bibinfo{person}{Ridha Hamila}, \bibinfo{person}{Ala Gouissem}, {and} \bibinfo{person}{Unal Devrim}.} \bibinfo{year}{2024}\natexlab{}.
\newblock \showarticletitle{Large Language Models (LLM) in Industry: A Survey of Applications, Challenges, and Trends}. In \bibinfo{booktitle}{\emph{2024 IEEE 21st International Conference on Smart Communities: Improving Quality of Life using AI, Robotics and IoT (HONET)}}. IEEE, \bibinfo{pages}{229--234}.
\newblock


\bibitem[Chu et~al\mbox{.}(2024)]%
        {10763608}
\bibfield{author}{\bibinfo{person}{Xiaoyu Chu}, \bibinfo{person}{Daniel Hofstätter}, \bibinfo{person}{Shashikant Ilager}, \bibinfo{person}{Sacheendra Talluri}, \bibinfo{person}{Duncan Kampert}, \bibinfo{person}{Damian Podareanu}, \bibinfo{person}{Dmitry Duplyakin}, \bibinfo{person}{Ivona Brandic}, {and} \bibinfo{person}{Alexandru Iosup}.} \bibinfo{year}{2024}\natexlab{}.
\newblock \showarticletitle{Generic and ML Workloads in an HPC Datacenter: Node Energy, Job Failures, and Node-Job Analysis}. In \bibinfo{booktitle}{\emph{2024 IEEE 30th International Conference on Parallel and Distributed Systems (ICPADS)}}. \bibinfo{pages}{710--719}.
\newblock
\href{https://doi.org/10.1109/ICPADS63350.2024.00097}{doi:\nolinkurl{10.1109/ICPADS63350.2024.00097}}


\bibitem[Chu et~al\mbox{.}(2025)]%
        {chu2025-llm-analysis}
\bibfield{author}{\bibinfo{person}{Xiaoyu Chu}, \bibinfo{person}{Sacheendra Talluri}, \bibinfo{person}{Qingxian Lu}, {and} \bibinfo{person}{Alexandru Iosup}.} \bibinfo{year}{2025}\natexlab{}.
\newblock \bibinfo{title}{An Empirical Characterization of Outages and Incidents in Public Services for Large Language Models}.
\newblock
\showeprint[arxiv]{2501.12469}~[cs.PF]
\urldef\tempurl%
\url{https://arxiv.org/abs/2501.12469}
\showURL{%
\tempurl}


\bibitem[Chu et~al\mbox{.}(2023)]%
        {2023-hotcloudperf-mlfailures}
\bibfield{author}{\bibinfo{person}{Xiaoyu Chu}, \bibinfo{person}{Sacheendra Talluri}, \bibinfo{person}{Laurens Versluis}, {and} \bibinfo{person}{Alexandru Iosup}.} \bibinfo{year}{2023}\natexlab{}.
\newblock \showarticletitle{How Do ML Jobs Fail in Datacenters? Analysis of a Long-Term Dataset from an HPC Cluster}. In \bibinfo{booktitle}{\emph{Proceedings of the International Conference on Performance Engineering, Coimbra, Portugal, April, 2023}}.
\newblock


\bibitem[Di et~al\mbox{.}(2019)]%
        {HPCfailures}
\bibfield{author}{\bibinfo{person}{Sheng Di}, \bibinfo{person}{Hanqi Guo}, \bibinfo{person}{Eric Pershey}, \bibinfo{person}{Marc Snir}, {and} \bibinfo{person}{Franck Cappello}.} \bibinfo{year}{2019}\natexlab{}.
\newblock \showarticletitle{Characterizing and Understanding HPC Job Failures Over The 2K-Day Life of IBM BlueGene/Q System}. In \bibinfo{booktitle}{\emph{2019 49th Annual IEEE/IFIP International Conference on Dependable Systems and Networks (DSN)}}. \bibinfo{pages}{473--484}.
\newblock
\href{https://doi.org/10.1109/DSN.2019.00055}{doi:\nolinkurl{10.1109/DSN.2019.00055}}


\bibitem[Gamell et~al\mbox{.}(2014)]%
        {gamell2014exploring}
\bibfield{author}{\bibinfo{person}{Marc Gamell}, \bibinfo{person}{Daniel~S Katz}, \bibinfo{person}{Hemanth Kolla}, \bibinfo{person}{Jacqueline Chen}, \bibinfo{person}{Scott Klasky}, {and} \bibinfo{person}{Manish Parashar}.} \bibinfo{year}{2014}\natexlab{}.
\newblock \showarticletitle{Exploring automatic, online failure recovery for scientific applications at extreme scales}. In \bibinfo{booktitle}{\emph{SC'14: Proceedings of the International Conference for High Performance Computing, Networking, Storage and Analysis}}. IEEE, \bibinfo{pages}{895--906}.
\newblock


\bibitem[Gupta et~al\mbox{.}(2017)]%
        {Gupta2017Nov}
\bibfield{author}{\bibinfo{person}{Saurabh Gupta}, \bibinfo{person}{Tirthak Patel}, \bibinfo{person}{Christian Engelmann}, {and} \bibinfo{person}{Devesh Tiwari}.} \bibinfo{year}{2017}\natexlab{}.
\newblock \showarticletitle{{Failures in large scale systems: long-term measurement, analysis, and implications}}.
\newblock In \bibinfo{booktitle}{\emph{{ACM Conferences}}}. \bibinfo{publisher}{Association for Computing Machinery}, \bibinfo{address}{New York, NY, USA}, \bibinfo{pages}{1--12}.
\newblock
\href{https://doi.org/10.1145/3126908.3126937}{doi:\nolinkurl{10.1145/3126908.3126937}}


\bibitem[Hu et~al\mbox{.}(2024)]%
        {Hu2024Characterization}
\bibfield{author}{\bibinfo{person}{Qinghao Hu}, \bibinfo{person}{Zhisheng Ye}, \bibinfo{person}{Zerui Wang}, \bibinfo{person}{Guoteng Wang}, \bibinfo{person}{Meng Zhang}, \bibinfo{person}{Qiaoling Chen}, \bibinfo{person}{Peng Sun}, \bibinfo{person}{Dahua Lin}, \bibinfo{person}{Xiaolin Wang}, \bibinfo{person}{Yingwei Luo}, \bibinfo{person}{Yonggang Wen}, {and} \bibinfo{person}{Tianwei Zhang}.} \bibinfo{year}{2024}\natexlab{}.
\newblock \showarticletitle{Characterization of Large Language Model Development in the Datacenter}. In \bibinfo{booktitle}{\emph{21st USENIX Symposium on Networked Systems Design and Implementation (NSDI 24)}}. \bibinfo{publisher}{USENIX Association}, \bibinfo{address}{Santa Clara, CA}, \bibinfo{pages}{709--729}.
\newblock
\showISBNx{978-1-939133-39-7}
\urldef\tempurl%
\url{https://www.usenix.org/conference/nsdi24/presentation/hu}
\showURL{%
\tempurl}


\bibitem[Isukapalli and Srirama(2024)]%
        {isukapalli2024systematic}
\bibfield{author}{\bibinfo{person}{Sucharitha Isukapalli} {and} \bibinfo{person}{Satish~Narayana Srirama}.} \bibinfo{year}{2024}\natexlab{}.
\newblock \showarticletitle{A systematic survey on fault-tolerant solutions for distributed data analytics: Taxonomy, comparison, and future directions}.
\newblock \bibinfo{journal}{\emph{Computer Science Review}}  \bibinfo{volume}{53} (\bibinfo{year}{2024}), \bibinfo{pages}{100660}.
\newblock


\bibitem[{\L}azuka et~al\mbox{.}(2024)]%
        {Lazuka2024Oct}
\bibfield{author}{\bibinfo{person}{Ma{\l}gorzata {\L}azuka}, \bibinfo{person}{Andreea Anghel}, {and} \bibinfo{person}{Thomas Parnell}.} \bibinfo{year}{2024}\natexlab{}.
\newblock \showarticletitle{{LLM-Pilot: Characterize and Optimize Performance of your LLM Inference Services}}.
\newblock \bibinfo{journal}{\emph{arXiv}} (\bibinfo{date}{Oct.} \bibinfo{year}{2024}).
\newblock
\href{https://doi.org/10.48550/arXiv.2410.02425}{doi:\nolinkurl{10.48550/arXiv.2410.02425}}
\showeprint{2410.02425}


\bibitem[Meduri(2024)]%
        {Meduri2024Oct}
\bibfield{author}{\bibinfo{person}{Sudeep Meduri}.} \bibinfo{year}{2024}\natexlab{}.
\newblock \showarticletitle{{Revolutionizing Customer Service : The Impact of Large Language Models on Chatbot Performance}}.
\newblock \bibinfo{journal}{\emph{International Journal of Scientific Research in Computer Science, Engineering and Information Technology}}  \bibinfo{volume}{10} (\bibinfo{date}{Oct.} \bibinfo{year}{2024}), \bibinfo{pages}{721--730}.
\newblock
\href{https://doi.org/10.32628/CSEIT241051057}{doi:\nolinkurl{10.32628/CSEIT241051057}}


\bibitem[@OpenAI(2024)]%
        {x_postopenai}
\bibfield{author}{\bibinfo{person}{@OpenAI}.} \bibinfo{year}{2024}\natexlab{}.
\newblock \bibinfo{title}{OpenAI's post on X apologizing for downtime}.
\newblock \bibinfo{howpublished}{\url{https://x.com/OpenAI/status/1867000372826607627}}.
\newblock
\newblock
\shownote{Accessed: 2025-01-15}.


\bibitem[Park et~al\mbox{.}(2023)]%
        {Park2023Jun}
\bibfield{author}{\bibinfo{person}{Ju-Won Park}, \bibinfo{person}{Xin Huang}, {and} \bibinfo{person}{Chul-Ho Lee}.} \bibinfo{year}{2023}\natexlab{}.
\newblock \showarticletitle{{Analyzing and predicting job failures from HPC system log}}.
\newblock \bibinfo{journal}{\emph{J. Supercomput.}} \bibinfo{volume}{80}, \bibinfo{number}{1} (\bibinfo{date}{June} \bibinfo{year}{2023}), \bibinfo{pages}{435--462}.
\newblock
\showISSN{0920-8542}
\href{https://doi.org/10.1007/s11227-023-05482-y}{doi:\nolinkurl{10.1007/s11227-023-05482-y}}


\bibitem[Relic(2023)]%
        {observability_report}
\bibfield{author}{\bibinfo{person}{New Relic}.} \bibinfo{year}{2023}\natexlab{}.
\newblock \bibinfo{title}{Observability Report}.
\newblock \bibinfo{howpublished}{\url{https://newrelic.com/resources/report/observability-forecast/2023/about-this-report}}.
\newblock
\newblock
\shownote{Accessed: 2025-01-15}.


\bibitem[Talluri et~al\mbox{.}(2021)]%
        {9659508}
\bibfield{author}{\bibinfo{person}{Sacheendra Talluri}, \bibinfo{person}{Leon Overweel}, \bibinfo{person}{Laurens Versluis}, \bibinfo{person}{Animesh Trivedi}, {and} \bibinfo{person}{Alexandru Iosup}.} \bibinfo{year}{2021}\natexlab{}.
\newblock \showarticletitle{Empirical Characterization of User Reports about Cloud Failures}. In \bibinfo{booktitle}{\emph{2021 IEEE International Conference on Autonomic Computing and Self-Organizing Systems (ACSOS)}}. \bibinfo{pages}{158--163}.
\newblock
\href{https://doi.org/10.1109/ACSOS52086.2021.00039}{doi:\nolinkurl{10.1109/ACSOS52086.2021.00039}}


\bibitem[Wang et~al\mbox{.}(2024)]%
        {Wang2024Jan}
\bibfield{author}{\bibinfo{person}{Yuxin Wang}, \bibinfo{person}{Yuhan Chen}, \bibinfo{person}{Zeyu Li}, \bibinfo{person}{Xueze Kang}, \bibinfo{person}{Zhenheng Tang}, \bibinfo{person}{Xin He}, \bibinfo{person}{Rui Guo}, \bibinfo{person}{Xin Wang}, \bibinfo{person}{Qiang Wang}, \bibinfo{person}{Amelie~Chi Zhou}, {and} \bibinfo{person}{Xiaowen Chu}.} \bibinfo{year}{2024}\natexlab{}.
\newblock \showarticletitle{{BurstGPT: A Real-world Workload Dataset to Optimize LLM Serving Systems}}.
\newblock \bibinfo{journal}{\emph{arXiv}} (\bibinfo{date}{Jan.} \bibinfo{year}{2024}).
\newblock
\href{https://doi.org/10.48550/arXiv.2401.17644}{doi:\nolinkurl{10.48550/arXiv.2401.17644}}
\showeprint{2401.17644}


\bibitem[White et~al\mbox{.}(2023)]%
        {white2023prompt}
\bibfield{author}{\bibinfo{person}{Jules White}, \bibinfo{person}{Quchen Fu}, \bibinfo{person}{Sam Hays}, \bibinfo{person}{Michael Sandborn}, \bibinfo{person}{Carlos Olea}, \bibinfo{person}{Henry Gilbert}, \bibinfo{person}{Ashraf Elnashar}, \bibinfo{person}{Jesse Spencer-Smith}, {and} \bibinfo{person}{Douglas~C Schmidt}.} \bibinfo{year}{2023}\natexlab{}.
\newblock \showarticletitle{A prompt pattern catalog to enhance prompt engineering with chatgpt}.
\newblock \bibinfo{journal}{\emph{arXiv preprint arXiv:2302.11382}} (\bibinfo{year}{2023}).
\newblock


\bibitem[Xu et~al\mbox{.}(2024)]%
        {Xu2024Dec}
\bibfield{author}{\bibinfo{person}{Weichao Xu}, \bibinfo{person}{Huaxin Pei}, \bibinfo{person}{Jingxuan Yang}, \bibinfo{person}{Yuchen Shi}, \bibinfo{person}{Yi Zhang}, {and} \bibinfo{person}{Qianchuan Zhao}.} \bibinfo{year}{2024}\natexlab{}.
\newblock \showarticletitle{{Exploring Critical Testing Scenarios for Decision-Making Policies: An LLM Approach}}.
\newblock \bibinfo{journal}{\emph{arXiv}} (\bibinfo{date}{Dec.} \bibinfo{year}{2024}).
\newblock
\href{https://doi.org/10.48550/arXiv.2412.06684}{doi:\nolinkurl{10.48550/arXiv.2412.06684}}
\showeprint{2412.06684}


\bibitem[Yigitbasi et~al\mbox{.}(2010)]%
        {Yigitbasi2010Oct}
\bibfield{author}{\bibinfo{person}{Nezih Yigitbasi}, \bibinfo{person}{Matthieu Gallet}, \bibinfo{person}{Derrick Kondo}, \bibinfo{person}{Alexandru Iosup}, {and} \bibinfo{person}{D. Epema}.} \bibinfo{year}{2010}\natexlab{}.
\newblock \showarticletitle{{Analysis and Modeling of Time-Correlated Failures in Large-Scale Distributed Systems}}.
\newblock \bibinfo{journal}{\emph{Telecommunications Policy - TELECOMMUN POLICY}} (\bibinfo{date}{Oct.} \bibinfo{year}{2010}), \bibinfo{pages}{65--72}.
\newblock
\href{https://doi.org/10.1109/GRID.2010.5697961}{doi:\nolinkurl{10.1109/GRID.2010.5697961}}


\bibitem[Yu et~al\mbox{.}(2024)]%
        {Yu2024Jun}
\bibfield{author}{\bibinfo{person}{Guangba Yu}, \bibinfo{person}{Gou Tan}, \bibinfo{person}{Haojia Huang}, \bibinfo{person}{Zhenyu Zhang}, \bibinfo{person}{Pengfei Chen}, \bibinfo{person}{Roberto Natella}, {and} \bibinfo{person}{Zibin Zheng}.} \bibinfo{year}{2024}\natexlab{}.
\newblock \showarticletitle{{A Survey on Failure Analysis and Fault Injection in AI Systems}}.
\newblock \bibinfo{journal}{\emph{arXiv}} (\bibinfo{date}{June} \bibinfo{year}{2024}).
\newblock
\href{https://doi.org/10.48550/arXiv.2407.00125}{doi:\nolinkurl{10.48550/arXiv.2407.00125}}
\showeprint{2407.00125}


\bibitem[Zhou et~al\mbox{.}(2024)]%
        {Zhou2024Jan}
\bibfield{author}{\bibinfo{person}{Pengyuan Zhou}, \bibinfo{person}{Lin Wang}, \bibinfo{person}{Zhi Liu}, \bibinfo{person}{Yanbin Hao}, \bibinfo{person}{Pan Hui}, \bibinfo{person}{Sasu Tarkoma}, {and} \bibinfo{person}{Jussi Kangasharju}.} \bibinfo{year}{2024}\natexlab{}.
\newblock \showarticletitle{{A Survey on Generative AI and LLM for Video Generation, Understanding, and Streaming}}.
\newblock \bibinfo{journal}{\emph{arXiv}} (\bibinfo{date}{Jan.} \bibinfo{year}{2024}).
\newblock
\href{https://doi.org/10.48550/arXiv.2404.16038}{doi:\nolinkurl{10.48550/arXiv.2404.16038}}
\showeprint{2404.16038}


\end{thebibliography}

\appendix

\newpage

\section{Screenshots from \textit{FAILS}}
\begin{figure}[h]
  \centering
  \includegraphics[trim={9.2cm 8cm 1cm 1cm}, clip, width=0.99\linewidth]{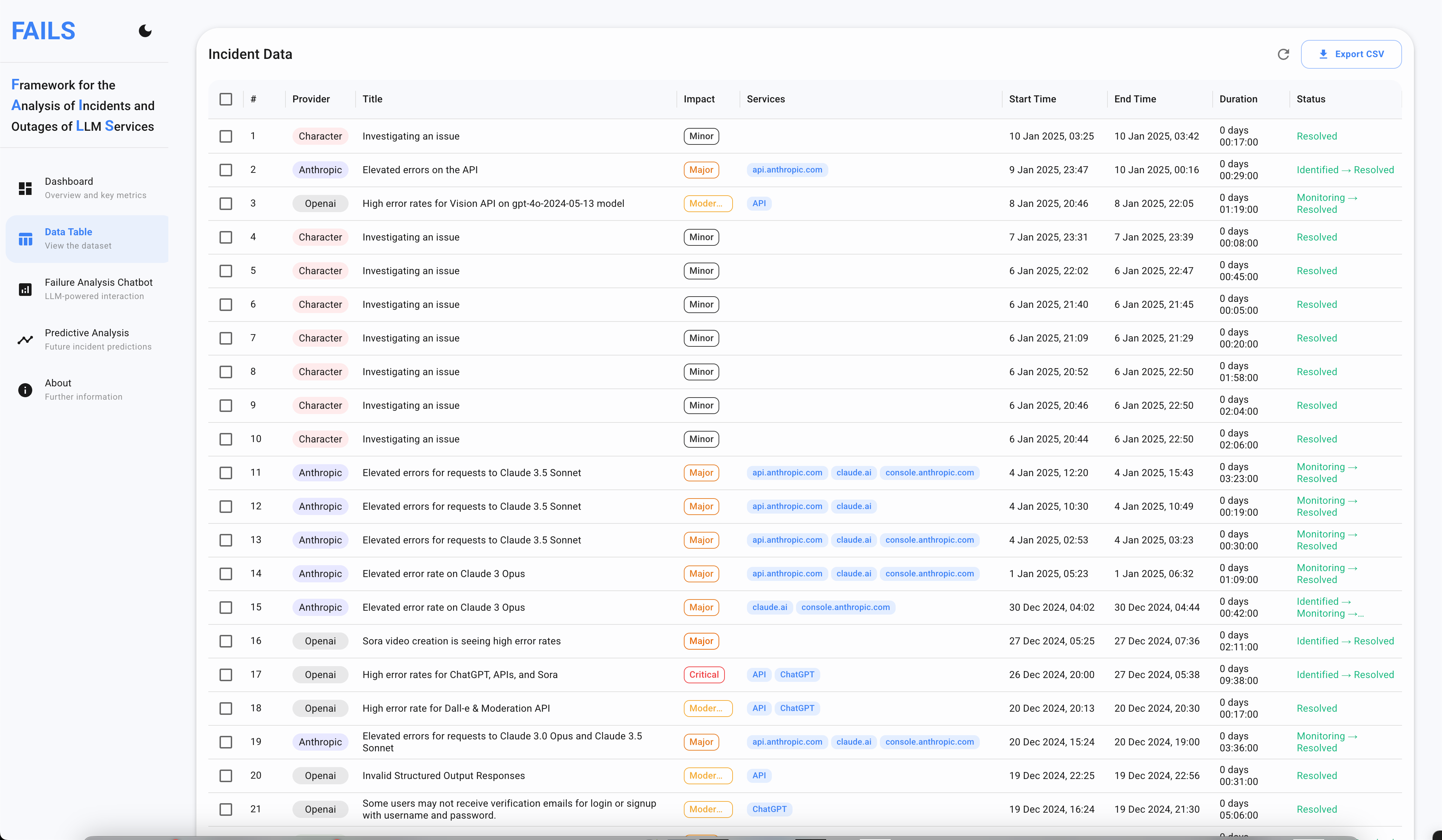}
  \vspace{-0.2cm}
        \caption{\textbf{The tabular dataset view in \textit{FAILS}.} It allows researchers to view each incident in a sorted format.}
       \label{fig:datatable}
\end{figure}

\begin{figure}[h]
  \centering
  \includegraphics[trim={0.8cm 0cm 1cm 3.5cm}, clip, width=0.99\linewidth]{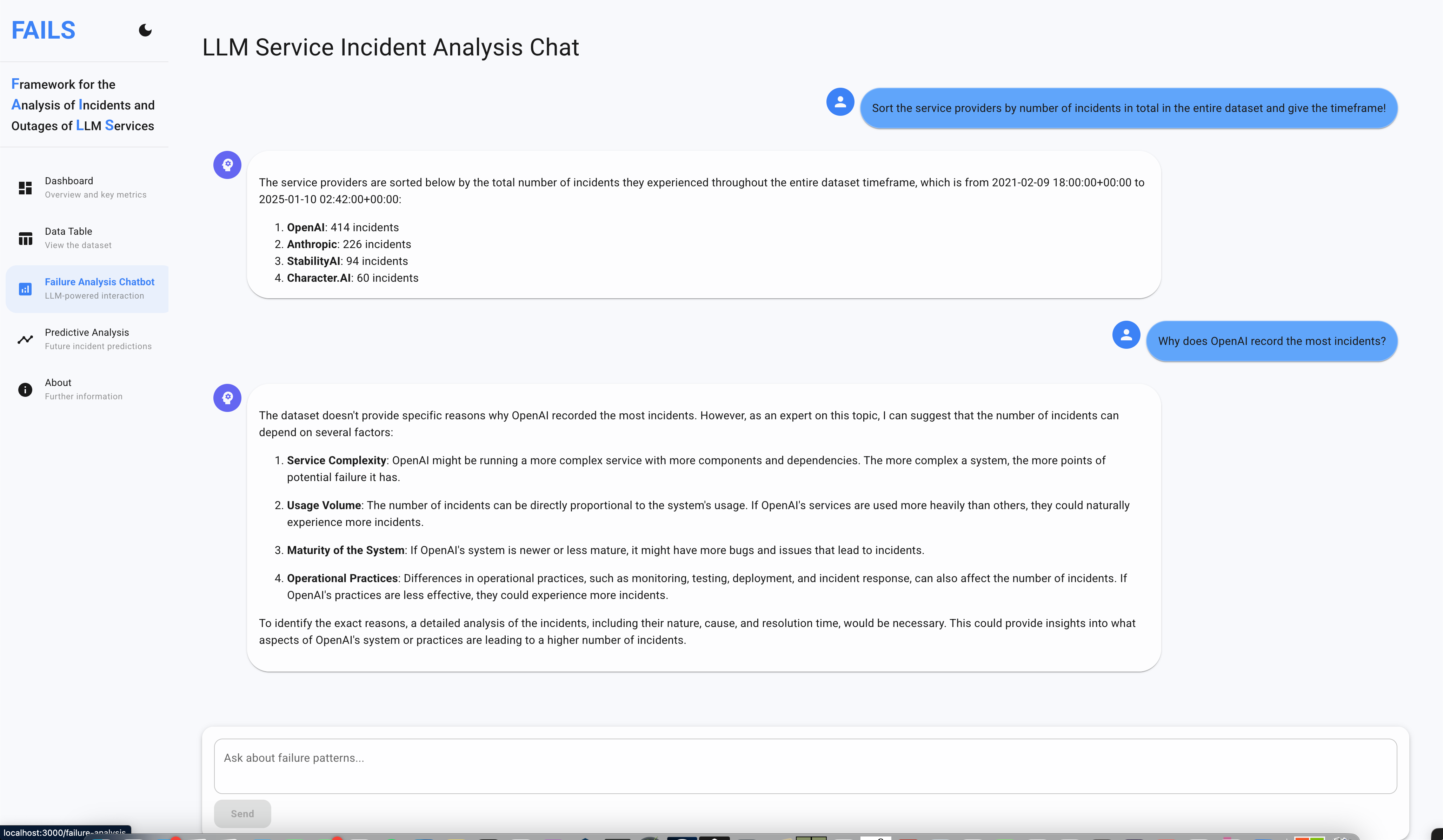}
  \vspace{-0.2cm}
        \caption{\textbf{The chatbot in \textit{FAILS}.} Here users can interact with the dataset in natural language.}
       \label{fig:analysisllm}
\end{figure}

\Cref{fig:datatable} shows the tabular data view (\bc{6a} in \Cref{fig:architecture}) and \Cref{fig:analysisllm} shows the long context interaction with the dataset through LLM chatbot (\bc{6b} in \Cref{fig:architecture}).

% Full screeenshots are accessible on \url{https://github.com/atlarge-research/FAILS}.
% Figures \ref{fig:mainpage}, \ref{fig:datatable}, \ref{fig:chatbot} and \ref{fig:analysisllm} show screenshots of subpages of the current version of \textit{FAILS}.

% \begin{figure*}[t]
%   \centering
%   \includegraphics[trim={0 14.3cm 0 0}, clip, width=0.99\linewidth]{figures/screenshots/mainpage2.png}
%   \vspace{-0.2cm}
%        \caption{\textbf{Main page from \textit{FAILS}.} This main dashboard page enables the user to select a date range and service, and generate plots based on the selection.}
%        \label{fig:mainpage}
% \end{figure*}

% \begin{figure*}[t]
%   \centering
%   \includegraphics[trim={13cm 8cm 12cm 8cm},clip,width=0.99\linewidth]{figures/screenshots/llmanalysis.png}
%   \vspace{-0.2cm}
%         \caption{\textbf{The LLM based image analysis in \textit{FAILS}.} The image is sent with an engineered prompt to an LLM service for analysis and a short text response is shown in a window.}
%        \label{fig:analysisllm}
% \end{figure*}

% \clearpage

% \twocolumn

\newpage

\section{List of Plots}\label{sec:plot-list}

Table \ref{tab:plot_overview} summarizes all the plots that are currently functional in \textit{FAILS} (\bc{4} and \bc{7a} in \Cref{fig:architecture}).
\begin{table}[h]
% \caption{\textbf{Comprehensive plot overview} of all visualizations in the dashboard section of \textit{FAILS}.}
\caption{The complete list of failure analysis types that \textit{FAILS} can perform and visualize.}
\label{tab:plot_overview}
\centering
\renewcommand{\arraystretch}{1.3}
\small
\begin{tabular}{cp{0.3\linewidth}p{0.55\linewidth}}
\toprule
\textbf{\#} & \textbf{Plot Name} & \textbf{Description} \\
\midrule
1  & Weekly Overview & Incident distribution across services by day of the week. \\
2  & Hourly Overview & Incident distribution across services by hour of the day (UTC).\\
3  & MTTR Distribution & Recovery time duration distributions for each service. \\
4  & MTTR by Provider & Recovery time duration grouped by provider. \\
5  & MTTR Boxplot & Boxplots showing MTTR statistics per service. \\
6  & MTBF Distribution & Time-between-failures distributions for each service. \\
7  & MTBF by Provider & Time-between-failures grouped by provider. \\
8  & MTBF Boxplot & Boxplots showing MTBF statistics per service. \\
9  & Resolution Activities & Resolution process duration and distribution. \\
10 & Status Combinations & Distribution of status combinations. \\
11 & Daily Availability & Day-by-day service uptime analysis. \\
12 & Service Co-occurrence & Correlation matrix of co-failing services. \\
13 & Co-occurrence Probability & Probability matrix of concurrent failures. \\
14 & Service Incidents & Breakdown of incidents by service. \\
15 & Incident Outage Timeline & Timeline of service disruptions. \\
16 & Autocorrelations & Time-series analysis of incident correlations. \\
17 & Incident Impact Distribution & Breakdown of incident severity by provider. \\
\bottomrule
\end{tabular}
\end{table}

\end{document}